\documentclass[twocolumn]{aastex62}

\usepackage{graphicx}	
\usepackage{amsmath}	
\usepackage{amssymb}	
\usepackage{threeparttablex}
\usepackage{comment}
\usepackage{bm}
\usepackage{nameref}
\usepackage{esint}

\newcommand{\bfv}{\bm{v}}
\newcommand{\bfb}{\bm{B}}
\newcommand{\bff}{\bm{F}}
\newcommand{\bnab}{\bm{\nabla}}
\newcommand{\bfsig}{\bm{\sigma}}
\newcommand{\mao}{\mathcal{M}_{A0}}
\newcommand{\dcr}{D_{\rm c}}
\newcommand{\dcrit}{D_{\rm crit}}
\newcommand{\ds}{D_{\rm s}}

\newcommand{\Pm}{{\rm Pm}}
\newcommand{\Pmcr}{{\rm Pm_{\rm c}}}
\newcommand{\M}{\mathcal{M}}
\newcommand{\e}{\'e}

\submitjournal{ApJ}

\shorttitle{Cosmic Ray Fluid - Plasma Coupling}
\shortauthors{Sampson et al.}
\begin{document}


\title{Cosmic ray and plasma coupling for isothermal supersonic turbulence in the magnetized interstellar medium}

\correspondingauthor{Matt L. Sampson}
\email{matt.sampson@princeton.edu}

\author[0000-0001-5748-5393]{Matt L. Sampson}
\affil{Department of Astrophysical Sciences, Princeton University,  Princeton, NJ 08544, USA}

\author[0000-0001-9199-7771]{James R. Beattie}
\affiliation{Department of Astrophysical Sciences, Princeton University, Princeton, NJ 08544, USA}
\affiliation{Canadian Institute for Theoretical Astrophysics, University of Toronto, 60 St. George Street, Toronto, ON M5S 3H8, Canada}

\author[0000-0001-7689-0933]{Romain Teyssier}
\affiliation{Department of Astrophysical Sciences, Princeton University, Princeton, NJ 08544, USA}

\author[0009-0009-2144-3912]{Philipp Kempski}
\affil{Department of Astrophysical Sciences, Princeton University,  Princeton, NJ 08544, USA}

\author[0000-0001-8558-5009]{Eric R. Moseley}
\affiliation{Department of Astrophysical Sciences, Princeton University,  Princeton, NJ 08544, USA}
\affiliation{Kavli Institute for Particle Astrophysics \& Cosmology (KIPAC), Stanford University, Stanford, CA 94305, USA}

\author[0000-0003-2407-1025]{Benoit Commer\c{c}on}
\affil{Univ Lyon, Ens de Lyon, Univ Lyon 1, CNRS, Centre de Recherche Astrophysique de Lyon UMR5574, 69007, Lyon, France} 

\author[0000-0003-0225-6387]{Yohan Dubois}
\affil{Institut d’Astrophysique de Paris, Sorbonne Universit\e, CNRS, UMR 7095, 98 bis bd Arago, 75014, Paris, France} 

\author[0000-0002-7534-8314]{Joakim Rosdahl}
\affil{Universite Claude Bernard Lyon 1, CRAL UMR5574, ENS de Lyon, CNRS, Villeurbanne, F-69622,  France} 

\begin{abstract}
    Cosmic rays (CRs) are an integral part of the non-thermal pressure budget in the interstellar medium (ISM) and are the leading-order ionization mechanism in cold molecular clouds. We study the impacts that different microphysical CR diffusion coefficients and streaming speeds have on the evolution of isothermal, magnetized, turbulent plasmas, relevant to the cold ISM. We utilized a two-moment CR magnetohydrodynamic (CRMHD) model, allowing us to dynamically evolve both CR energy and flux densities with contributions from Alfv\'enic streaming and anisotropic diffusion. We identify \emph{coupled} and \emph{decoupled} regimes, and define dimensionless Prandtl numbers $\Pmcr$ and $\Pm_{\rm s}$, which quantify whether the plasma falls within these two regimes. In the coupled regime -- characteristic of slow streaming ($\Pm_{\rm s} < 1$) and low diffusion ($\Pmcr < 1$) -- the CR fluid imprints upon the plasma a mixed equation of state between $P_{\rm{c}} \propto \rho^{4/3}$ (relativistic fluid) and $P_{\rm{c}} \propto \rho^{2/3}$ (streaming), where $P_{\rm{c}}$ is the CR pressure, and $\rho$ is the plasma density. By modifying the sound speed, the coupling reduces the turbulent Mach number, and hence the amplitude of the density fluctuations, whilst supporting secular heating of the CR fluid. In contrast, in the decoupled regime ($\Pm_{\rm s} > 1$ or $\Pmcr > 1$) the CR fluid and the plasma have negligible interactions. We further show that CR heating is enabled by coherent structures within the compressible velocity field, with no impact on the turbulence spectrum of incompressible modes.
\end{abstract}


\section{Introduction} 
\label{sec:intro}
The study of cosmic rays (CRs) has gained significant attention over the last decade \citep[see e.g.,][for a recent, comprehensive review]{Ruszkowski2023_cr_review} due to the impact they have on ionization inside molecular clouds \citep{padovani2009cosmic}, for their vertical pressure support and ability to drive galactic winds in disks \citep{uhlig2012, booth2013, hanasz2013,recchia16,simpson2016role, wiener2017, pfrommer17,mao2018galactic, Dashyan2020, crocker2021cosmic, girichidis22, thomas23,montero24}, as well as other impacts on the thermodynamics within galactic plasma environments \citep{zweibel2017basis}. 

For $\sim$GeV energy CRs evolving in $\sim$$\mu \rm{G}$ magnetic fields, which represents the largest number density of the CR population in the Milky Way Galaxy, the gyroradius is of order $10^{-6} \ \rm{pc}$ \citep{zweibel2013microphysics}. For these types of CRs, on scales of order a parsec and above \citep{wentzel74, garcia1987cosmic, yan2004cosmic}, CR propagation is effectively diffusive \citep{Skilling71a, zweibel2013microphysics, Krumholz2020CosmicGalaxies, sampson_2022_turb}. This could be due to pitch-angle scattering of CRs along magnetic field lines on gyroradii scales, whereby CRs undergo a relativistic random walk around field lines \citep{Skilling71a,skilling1975cosmic,goldstein1976,bieber1988,shalchi2004nonlinear,shalchi2009analytical}. We call the diffusion coefficient associated with this random walk the \textit{microphysical} diffusion coefficient, noted here $D_c$. 

This microphysical diffusion is likely due to the presence of Alfv\'en waves of wavelength comparable to the CR gyroradius. These waves can be excited by the CR streaming instability (CRSI) \citep{Lerche67a,Kulsrud69a}, and are amplified until there is a balance between wave growth, through resonant particle - wave interactions, and wave dampening from the ISM plasma (with different dampening mechanisms in different phases; \citealt{Xu2022_streaming}). The end result is that the resonant CR population is characterized by a net drift velocity which can be as low as the ion Alfv\'en speed $v_{A,\rm{ion}} = B/\sqrt{\chi \rho}$ of the plasma, where $B$ is the local magnetic field strength, $\chi$ is the mass ionization fraction, and $\rho$ is the local plasma density \citep[][and references therein]{Kulsrud69a,Farmer04a,zweibel2013microphysics}.   

Diffusive transport of CRs can also be realized due to (time-dependent) magnetic structures on even larger macroscopic scales, comparable to the outer scale of the turbulence. If either the microphysical diffusion timescale $\ell_0^2/D_c$ or the streaming timescale $\ell_0 / \bfv_{A}$ (where $\ell_0$ is the outer scale of the turbulence) are comparable to the dynamical timescale of the plasma turbulence, the CRs and the plasma are coupled on the scales of the turbulence (scales well above the gyroradii, and greater than the mean free path $\lambda_{\rm mfp}$ due to pitch-angle scattering). If this happens, a new diffusive transport mechanism emerges, which is due to both the complex and chaotic structural geometry of the field lines themselves and their time dependence \citep{Krumholz2020CosmicGalaxies,Yuen2020_curvature,Beattie2022_va_fluctuations,sampson_2022_turb,kempski23,kempski24,Kriel2025_curvature_and_dynamo}. When we coarse-grain the CR transport on scales of order the magnetic correlation lengths ($\sim 100\,\rm{pc}$ in the Milky Way; \citealt{Haverkorn2015_MW_magnetic_field_review}), this acts as an additional effective diffusion coefficient, describing the diffusion of a CR population as it moves through multiple correlation scales of the stochastic field lines. 

We call the diffusion on these scales the \textit{macrophysical} diffusion, which was studied in detail in \citet{Beattie2022_va_fluctuations} and \citet{sampson_2022_turb}, which include a CR streaming model, but where the microphysical diffusion coefficient was set to zero, and the measured diffusion coefficients were only a function of the background plasma turbulence. These two diffusion mechanisms are vastly different in both their underlying physical mechanisms (pitch-angle scattering versus turbulent advection/field-line random walk) and the length scales ($\lambda_{\rm mfp}=1 \,\rm{pc}$ compared to $\sim 100\,\rm{pc}$) at which they dominate. As long as the plasma and CRs are dynamically coupled, the observed CR diffusion coefficients, $D_c \sim 10^{26-30}\,\rm{cm}^2\,\rm{s}^{-1}$ \citep{jokipii1969cosmic,cesarsky80,Gabici2010_observational_D_W28,Heesen2023_D_observed_M51, wiener2017, hanasz2021simulations} are going to be a combination of both micro and macrophysical processes.  

It is becoming common to include a dynamical CR model in magnetohydrodynamic (MHD) fluid simulations that probe the evolution of galaxies \citep{jubelgas2008,wiener2017, thomas_19,armilotta2021, werhahn2021,thomas21, hopkins2021cosmic, werhahn2021,hopkins_2022_cR_model, kempski, farcy2022, MartinAlvarez2023_cr_galaxy_models} as well as studies directly probing CR transport in turbulent fields \citep{reichherzer2020turbulence, reichherzer22, laza21, kempski23, laza23, lemoine23}. These MHD simulations typically are limited to resolve scales well above the gyroradii of GeV CRs, and hence are not able to capture microphysical CR processes \citep{kulsrud2005plasma,thomas22}. 

Instead, one can adopt a relativistic, non-thermal fluid approach to model CR transport (as is commonly done in the context of radiation transport). In this \textit{CRMHD} framework the CR fluid is coupled to and evolved alongside the MHD equations for the thermal plasma. As with any radiation transport scheme, CR fluid models have closure schemes that truncate the set of equations one has to solve coupled to the plasma. 

The one-moment closure for a purely diffusive CR fluid is $\bm{F}_{\rm{c}} = \bfv(E_{\rm{c}} + P_{\rm{c}}) - \dcr \nabla E_{\rm{c}}$, where $\bfv$ is the fluid velocity, $\bm{F}_{\rm{c}}$ is the CR (lab-frame) flux density, and $E_{\rm{c}}$ is the CR energy density and $\dcr$ is the microphysical diffusion coefficient \citep{salem2014cosmological,dubois16, commercon19}. $E_{\rm{c}}$ is then co-evolved with the plasma, and $\bm{F}_{\rm{c}}$ is a prescribed quantity, completely in equilibrium with the advection and pressure gradients \citep{oneMom,commercon19,thomas22}. These methods have numerical stability issues at $|\nabla E_{\rm{c}}| = 0$, which require artificial diffusion or regularization, and due to the parabolic nature of the one-moment equation, have prohibitive time-stepping requirements. Furthermore, $\bm{F}_\mathrm{c}$ for streaming and diffusion has to be prescribed, rather than dynamically evolved \citep{sharma2009}. More specifically, the problem is for the streaming term which indeed produces severe constraints on the time step for explicit methods because of $D_{\rm stream}\propto 1/\nabla E_{\rm{c}}$. This can be addressed with implicit solvers such as in \citet{Dubois2019}.

For the two-moment method (analogous to the M1 closure in radiative transport) $|\partial_t \bm{F}_{\rm{c}}| \neq 0$ and $\bff_{\rm{c}}$ is evolved together with the plasma, resulting in a set of hyperbolic, causal CR evolution equations \citep{2mom,Hopkins2022_two_moment_cRMHD_RSOL,thomas22}. This results in numerically stable equations that can utilize explicit integration methods limited by only the CFL condition, preserving physical propagation speeds in regions with $|\nabla E_{\rm{c}}| \approx 0$, where $\bm{F}_{\rm{c}}$ is not in equilibrium \citep{2mom, thomas_19,thomas21}, although this method generally requires to severely reduce the speed of light to an ad hoc speed to be useful in practice. The $\bm{F}_{\rm{c}}$ equation that we use in this study is based on \citet{2mom}, which includes flux contributions from CR advection, anisotropic diffusion, and Alfv\'enic streaming, allowing us to probe the details of CRMHD turbulence in ISM-type conditions in a numerically stable manner. 

An important question to understand in CRMHD is what effect does the CR fluid have on the evolution of the plasma itself, the so-called backreaction (or what we call coupling throughout this study) \citep{caprioli2009, orlando2012, zweibel2013microphysics}. \citet{commercon19} explore this question utilizing a one-moment CRMHD fluid in a series of 3D turbulence simulations, where the anisotropic microphysical diffusion coefficient, driving scale, magnetization and turbulence was varied. They define a theoretical \emph{critical} (turbulent) diffusion limit 
\begin{align}\label{eq:turbulent_diffusion}
    \dcrit = \sigma_v\ell_0  \sim 3.1 \times 10^{25} \left(\frac{\ell_0}{1 \rm{pc}} \right)\; \rm{cm}^2\,\rm{s}^{-1},
\end{align}
where $\sigma_v$ is the turbulence velocity and $\ell_0$ is the turbulence coherence scale, which defines a critical diffusion coefficient from the background turbulence in the velocity field i.e., the effective diffusion coefficient of the turbulent velocity field. In their experiments they found that for values less than $\dcr \sim 10^{25}\; \rm{cm^2\; s^{-1}} \approx \dcrit$ the CR fluid dynamically coupled to the plasma, acting to reduce the density variance of the plasma density with decreasing $\dcr$. However, this study uses a one-moment CR-fluid model that neglects the CRSI. 

In a follow-up paper by~\citet{Dubois2019}, it was shown that in the absence of diffusion but in presence of the streaming instability, CRs are efficiently redistributed in the ISM. \citet{bustard23} explore the effect of a CR fluid on the energy cascade of compressible CRMHD turbulence using a two-moment scheme. They find the CR fluid is able to dampen large-scale motions in the plasma in the streaming-dominated regime, reducing the kinetic energy of the plasma by up to an order of magnitude without affecting the turbulent cascade. \citet{bustard23} also note that in the diffusion-dominated limit of CR transport, there is a preferential dampening of compressible modes in the turbulence, resulting in predominantly solenoidal, incompressible turbulence on smaller scales in their simulations with differences to the overall structure of the kinetic energy spectra. This work considers strictly subsonic turbulence, with fully curl-free driving (allowing solenoidal modes to arise naturally from compressible mode interactions).

In our study, we perform a similar analysis to \citet{commercon19}, aiming to probe the nature of the CR-plasma coupling via a range of statistics. However, we use a two-moment CR transport model, with the same flux equation as in \citet{bustard23}, originally from \citet{2mom}, where we evolve both the CR energy and flux density in a self-consistent, hyperbolic manner. We include flux contributions from advection, anisotropic diffusion, and streaming using an M1-like approximation for the second-order closure conditions from by Rosdahl et al.~(in prep.). We perform a suite of CRMHD simulations using a modified version of the \texttt{RAMSES} code \citep{RAMSES} to study the effect of different streaming parameters and microphysical diffusion coefficients on the pressure, CR heating and turbulent spectra in CRMHD. 

The structure of this paper is as follows; in \autoref{sec:methods} we discuss the CRMHD fluid model and implementation, and describe the suite of simulations used in this study. Critically, we define three new dimensionless parameters that describe when the CR fluid and MHD plasma should be dynamically coupled, and then further which term in the CR flux equation should be the strongest in the coupling. In \autoref{sec:pdfs} we characterize the turbulent fields, focusing on how the additional CR fluid can modify the equation of state of the plasma. In \autoref{sec:corr_scale} we measure spectrum and correlation scales for a variety of plasma and CR fluid variables.  In \autoref{sec:heating} we show how the CR fluid is able to be secularly heated (CRs, secularly reaccelerated) by the turbulence in the coupled regime. Finally, in \autoref{sec:conclusion} we list the key results and limitations of the study. We show a table of the definitions of our variables used throughout the study in \autoref{tab:notation}.
\begin{table*}[ht]
\centering
\caption{Definitions of variables}
\begin{tabular}{>{}lp{6cm}p{8.0cm}}
\hline \hline
Parameter & Definition/equation  & Notes \\
\hline
$L$ & Computational domain size & 1 in simulation units \\
$\ell_0$ & Outer-scale of turbulence & $\approx L/2$, measured directly from power spectra \\
$\sigma_v$ & $\sqrt{1/N\sum_j\sum_i(v_{j,i} - \bar{v}_{j})^2}, \;j \in \{x,y,z \}$ & rest-frame velocity dispersion, where $i$ is the grid-cell index and $N$ is the total number of grid cells. \\
$c_s$ & $\sqrt{\gamma P_{\rm therm} / \rho}, \quad\quad \gamma = 1$& isothermal plasma sound speed, $c_s=1$ \\
$c_{\rm{eff}}$ & $\sqrt{\gamma_{\rm{c}}\frac{P_{\rm{c}}}{\rho} + 1}, \quad\quad \gamma_{\rm{c}} = 4/3$ & effective sound speed due to thermal gas and CRs, where $\gamma_{\rm{c}}$ is the adiabatic index of the CR fluid \\
$\tau$ & $\ell_0 / \sigma_v$ = $t_{\rm{turb}}$ & correlation time, and equivalently, outer-scale turbulent turnover time \\
$\mathcal{M}$ & $\sigma_v / c_s$ & sonic Mach number, $\approx 4$ for all simulations \\
$\mathcal{M}_{\rm{eff}}$ & $\sigma_v / c_{\rm{eff}}$ & effective sonic Mach number \\
$\bm{b}$ & $\bfb/B$& the magnetic field unit vector\\
$\bm{v}_{A0}$ & $B_0/\sqrt{\rho}$& mean field Alfv\'en speed\\
$\mathcal{M}_{A0}$ & $\sigma_v / v_{A0}$ & the mean field Aflv\'en Mach number\\
$\beta$ & $2\mao^2/\mathcal{M}^2$ & the mean field plasma beta for an isothermal equation of state\\
$D_{\rm c}$ & parallel component of \autoref{eq:diffusion_coefficient} & microphysical (sub-grid) parallel diffusion coefficient\\
$D_{\rm crit}$ &$\sigma_v \ell_0$ $(t_{\rm{turb}} = t_{\rm{diff}})$ & critical diffusion coefficient \citep{commercon19} \\
$\rm{Pm_{c}}$ & $D_{\rm c} / D_{\rm crit}$ & CR diffusion Prandtl number; see \autoref{eq:Pms}\\
$\rm{Pm_{s}}$ & $v_{A0}\ell_0 / D_{\rm crit}$& CR streaming Prandtl number; see \autoref{eq:Pms}\\
$\rm{Pm_{f}}$ & $\rm{Pm_{c}} / \rm{Pm_{s}}$ & see \autoref{eq:Pms}\\
$\ell_{v}/L$ &  correlation scale for velocity field &  see \autoref{sec:definition_corr} for full definition\\
$\ell_{B}/L$ & correlation scale for $\bm{B}$ field & \texttt{"} \\
$\ell_{\rho}/L$ &correlation scale for density  & \texttt{"}\\
$\ell_{F_{\rm{c}}}/L$  &correlation scale for CR flux& \texttt{"} \\
$Q_c$ &$- \left\langle P_{\rm{c}} \nabla \cdot (\bfv + \bfv_s) \right\rangle$ & CR fluid heating rate, see \autoref{eq:one-mom}\\
$s$ & $\log(\rho / \rho_0)$ & the logarithim of the mean normalized plasma density, $\rho/\rho_0$ \\
\hline \hline
\end{tabular}
\label{tab:notation}
\end{table*}

\section{Numerical Methods}
\label{sec:methods}
\subsection{CR transport model}
We use the two-moment CR fluid approach described in \citet{2mom}, implemented in \texttt{RAMSES} by Rosdahl et al.~(in prep.), where we treat CRs as a non-thermal, relativistic fluid with adiabatic index $\gamma_{\rm{c}}=4/3$. Following \citet{2mom} we solve the time-dependent equations for CR energy and flux density,
\begin{equation}
\label{eqn:energy}
    \frac{\partial E_{\rm{c}}}{\partial t} + \bnab \cdot \bff_{\rm{c}} = (\bfv + \bfv_s) \cdot (\bnab \cdot \mathbb{P}_c),
\end{equation}
\begin{equation}
\label{eqn:flux}
        \frac{1}{V^2_\mathrm{m}}\frac{\partial \bff_{\rm{c}}}{\partial t} + \bnab \cdot \mathbb{P}_c = -\bfsig_c (\gamma_c-1) \cdot \left[ \bff_{\rm{c}} - \bfv \cdot (\mathbb{E}_c + \mathbb{P}_c) \right],
\end{equation}
where $\mathbb{P}_c = \mathbb{E}_c/3$ is the standard closure for an isotropic, relativistic fluid and hence $\mathbb{E}_c = E_{\rm{c}}\mathbb{I}$, where $\mathbb{I}=\delta_{ij}$ is the identity tensor, $\bfv$ is the plasma velocity, $\gamma_c$ is the adiabatic index of the CR fluid (4/3), and $\bfv_s$ is the streaming speed, such that 
\begin{equation}
\label{eq:stream}
    \bfv_s = -v_A\frac{\bfb \cdot \bnab P_{\rm{c}}}{\|\bfb \cdot \bnab P_{\rm{c}}\|}\bm{b},
\end{equation}
i.e., the population of CRs stream at the Alfv\'en speed in the direction down the CR pressure gradient along the magnetic field $\bm{b}$. $V_{\rm{m}}$ is the CR flux propagation speed, and in practice parameterizes the reduced speed of light approximation \citep[e.g.,][]{Hopkins2022_two_moment_cRMHD_RSOL}, reducing the stiffness of the equations. \citet{2mom} show that as long as $V_m \gg \mathtt{max}\left\{\|\bfv_A + \bfv\|\right\}$ the results are insensitive to the specific value of $V_m$ chosen. Importantly, in \autoref{eqn:flux}, \citet{2mom} introduce a term $\bm{\sigma}_c$ which is defined as the interaction coefficient as,
\begin{equation}
\label{eq:interaction}
    \bfsig_c^{-1} = \underbrace{\mathbb{D}_c}_{\mathrm{diffusion}} +\  \overbrace{\frac{1}{\sqrt{\rho}}\frac{\bfb \otimes \bfb}{\|\bfb \cdot \bnab P_{\rm{c}}\|}  \cdot (\mathbb{E}_c + \mathbb{P}_c)}^{\mathrm{streaming}},
\end{equation}
where 
\begin{align}\label{eq:diffusion_coefficient}
    \mathbb{D}_c = D_{\rm{c},\perp} \mathbb{I} - ( D_{\rm{c},\perp} - D_{\rm{c},\parallel} ) \bm{b}\otimes\bm{b},
\end{align}
is the anisotropic diffusion tensor of the CRs, where $\bm{b} = \bm{B}/B$ is the unit vector of the magnetic field, $D_{\rm{cr},\perp}$ is the microphysical diffusion coefficient in the direction perpendicular to the local magnetic field, and $D_{\rm{cr},\parallel}$ is the parallel diffusion coefficient. The role of the interaction coefficient becomes even more apparent when we rewrite \autoref{eqn:flux} as
\begin{align}
    \frac{\partial \bm{F}_{\rm{c}}}{\partial t} = -\frac{\bm{F}_{\rm{c}} - \bm{F}_{c,\rm{eq}}}{\tau_{\rm eq}},
\end{align}
where $\tau_{\rm eq} = (\bm{\sigma}_c V_m^2)^{-1}$ is the equilibration time of the two-moment model into the one-moment model equilibrium,
\begin{align}\label{eq:equilibrium_flux}
    \bm{F}_{c,\rm{eq}} = \bm{v}(E_{\rm{c}} + P_{\rm{c}}) - \bm{\sigma}_c^{-1}\cdot \nabla \cdot \mathbb{P}_c.
\end{align}

\subsection{Turbulent CRMHD fluid model}
In our \texttt{RAMSES} implementation, similarly to \citet{2mom}, we couple the two-moment cosmic ray fluid to the ideal, compressible, isothermal magnetohydrodynamic fluid model. The master equations are 
\begin{subequations} \label{eq:CRMHD}
\begin{align}
\label{eq:CRMHD-cont}
&\frac{\partial \rho}{\partial t} + \bnab \cdot (\rho \bfv) = 0, \\
\label{eq:CRMHD-mom}
&\frac{\partial (\rho \bfv)}{\partial t} + \bnab \cdot (\rho \bfv\otimes\bfv- \bfb\otimes\bfb + \mathbb{P}) = \nonumber \\ 
&\hspace{35mm}\bm{\sigma}_c (\gamma_c-1)  \cdot [\bff_{\rm{c}} - \bfv \cdot (\mathbb{E}_c + \mathbb{P}_c)] \nonumber\\ &\hspace{35mm} + \rho\bm{f}, \\ 
\label{eq:CRMHD-erg}
&\frac{\partial E}{\partial t} + \bnab \cdot [(E + P)\bfv - \bfb(\bfb\cdot \bfv)] = \nonumber \\
&\hspace{35mm}-(\bfv + \bfv_s) \cdot (\bnab \cdot \mathbb{P}_c) +  Q_{\mathrm{cool}}, \\ 
\label{eq:CRMHD-ecr}
&\frac{\partial E_{\rm{c}}}{\partial t} + \bnab \cdot \bff_{\rm{c}} = 
(\bfv + \bfv_s) \cdot (\bnab \cdot \mathbb{P}_c) \\
\label{eq:CRMHD-fc}
&\frac{1}{V_m^2}\frac{\partial \bff_{\rm{c}}}{\partial t} + \bnab \cdot \mathbb{P}_c = -\bm{\sigma}_c (\gamma_c-1)  \cdot [\bff_{\rm{c}} - \bfv \cdot (\mathbb{E}_c + \mathbb{P}_c)],\\
&\frac{\partial \bfb}{\partial t} + \bnab\cdot (\bfv \otimes \bfb - \bfb \otimes \bfv) = 0, \\ 
& P = c_s^2\rho + B^2/2, \hspace{5mm} \bnab\cdot\bfb = 0,
\end{align}
\end{subequations}
where $\bm{f}$ is our stochastic forcing field further defined in \autoref{sec:driving}, $E$ is the total plasma energy (excluding CRs) containing the kinetic, magnetic, and internal energies respectively $E = E_{\rm kin} + E_{\rm mag} + E_{\rm therm}$, $Q_{\mathrm{cool}}$ is the cooling term for the plasma, which is automatically set to ensure an isothermal plasma at each timestep, $\rho$ is the plasma density, $c_s$ is the sound speed, and $P$ is the total plasma pressure (again excluding CRs), which has thermal $P_{\rm therm} = \rho c_s^2 $ and magnetic $P_{\rm mag} = B^2/2$ components. We set $V_m = 10^{-3}c \approx 560c_s \gg \mathtt{max}\left\{\|\bfv_A + \bfv\|\right\}$ where $c$ is the speed of light. We solve the equations using the MUSCL-Hancock scheme \citep{teyssier2006, fromang2006} and the HLLD approximate Riemann solver in a three-dimensional, triply-periodic volume, discretized on a $256^3$ grid. Based on the numerical dissipation estimates of \cite{Shivakumar2025_numerical_dissipation} for a similar 5 wave approximate Riemann solver, we believe the effective plasma and magnetic Reynolds numbers of our $256^3$ grid to be between $500$ and $1000$. 

\begin{figure}
    \centering
    \includegraphics[width=0.95\linewidth]{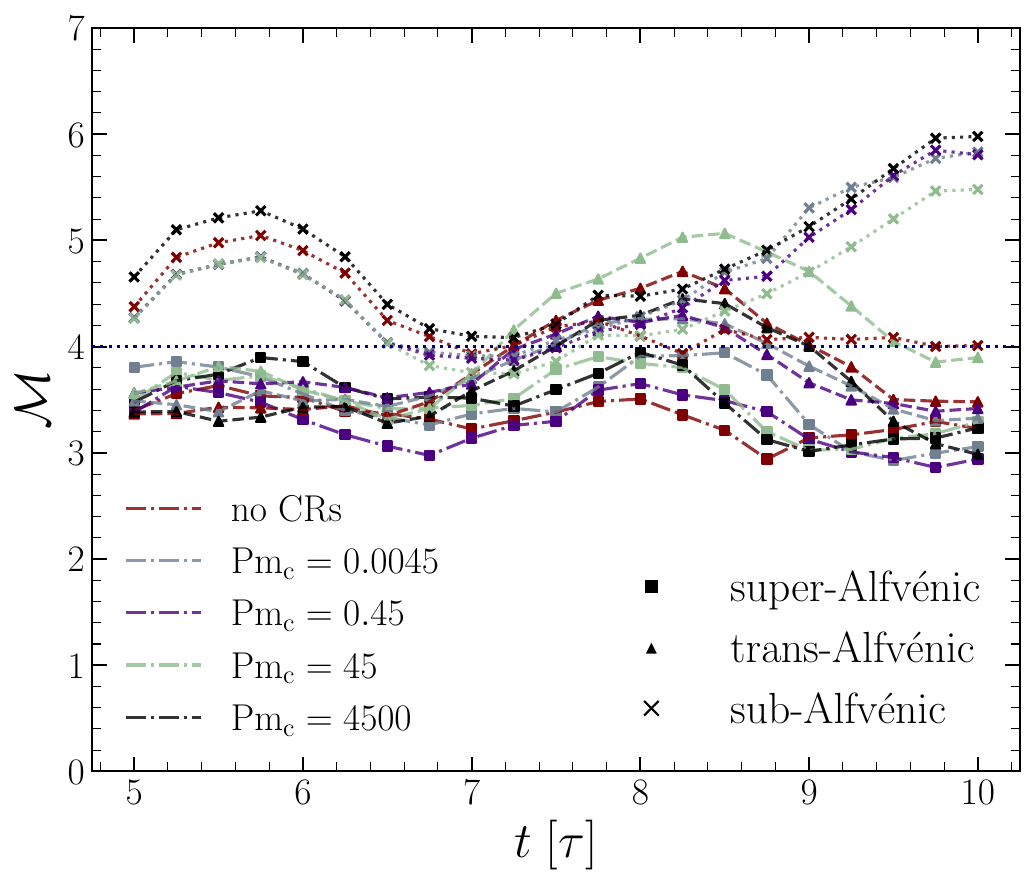}
    \caption{Turbulent Mach number, $\mathcal{M}$, plotted as a function of time, $t/\tau$, for all simulation runs. The $\mathcal{M}$ values are mass-weighted and averaged over volume. The different $\mao$ regimes are indicated by the markers, with the value of ratio between the microphysical diffusion coefficient and the turbulence diffusion coefficient $\Pm_{\rm cr}$ (see \autoref{eq:Pms}), indicated with the color. We show results for the last $5\tau$ of our simulation. This allows all simulations to reach a statistically stationary state \citep{Beattie2021_spdf} with $\mathcal{M} \approx 4$. All statistics in the study will be averaged over multiple realizations across the $5\leq t\leq10\tau$ interval shown in this plot. Note that we drive with exactly the same momentum flux injection, so there is some scatter between the simulations with different magnetic field strengths (Alfv\'enic state of the turbulence, shown with the marker). See \autoref{tab:trials} for a list all of the averaged $\mathcal{M}$.}
    \label{fig:mach}
\end{figure}

\subsubsection{Turbulent driving}
\label{sec:driving}
We drive turbulence in our plasma with a finite-correlation time generalized Ornstein-Uhlenbeck process \citep{eswaren_ornstein, schmidt2009numerical, federrath2010comparing} with a forcing field $\bm{f}$ satisfying the stochastic differential equation in $k$-space,
\begin{align}
    d\hat{\bm{f}}(\bm{k}, t) = -\tilde{\bm{f}}(\bm{k}, t)\frac{dt}{\tau}
    + F_0(\bm{k}) \bm{P}_\eta(\bm{k}) \cdot d\bm{W}_t.
\end{align}
where $\tilde{\bm{f}}$ is the Fourier transform of the acceleration field, $\bm{k}$ is the wavenumber, $\tau$ is the correlation timescale, which we set to equal the eddy turnover time on the outer scale (Strouhal number $=1$), $F_0$ is the mask for the driving modes in $k$ space, selecting only the driving modes we want to excite, $d\bm{W}_t$ is a Wiener process and $\bm{P}_{\eta}$ is the projection tensor, performing the Helmholtz decomposition on the forcing field, separating the compressible and solenoidal modes, parameterized by $\eta$ \citep{Federrath2010,brucy_driving}. The forcing field $\bm{f}(\bm{x},t)$ is defined by
\begin{equation}
\label{eq:driving}
    \bm{f}(\bm{x},t) = g(\eta) \bm{f}_{\mathrm{rms}} \int \bm{\tilde{f}}(\bm{k}, t) e^{i\bm{k}\cdot \bm{x}}d^3\bm{k},
\end{equation}
where $\eta$ is the energy fraction of compressible versus solenoidal driving modes, $\bm{f}_{\mathrm{rms}}$ is the force injected into the simulation, and $g(\eta)$ is an empirical correction function ensuring that the resulting integral over the Fourier modes is equal to $\bm{f}_{\mathrm{rms}}$, independent of the compressive fraction $\eta$ \citep{brucy_driving}. 

We choose $g$ and $\bm{f}_{\mathrm{rms}}$ to be exactly the same for each simulation, meaning that the injected energy flux is the same for all simulations. We set $\eta=0.5$ for an exact energy equipartition between compressive and solenoidal modes components in the driving and inject isotropically on the $k = 1 - 32$ modes, with $F_0(\bm{k})$ following a power-law scaling $F_0(|\bm{k}|) \propto k^{-2}$, mimicking \citet{Burgers1948_turbulence}-type turbulence (similar to \citealt{Nam2021_power_law_driving}). 

This provides an outer-scale for the turbulence $\ell_0 \approx L/2$ where $L$ is the size of our domain and $\ell_0 \propto \int k^{-1} F_{0}(k) dk$ is the outer scale of the velocity field. We verify that this is the same as the correlation scale in the turbulence through direct measurements of the isotropic correlation length of the velocity field in \autoref{sec:corr_scale}, tabulated in \autoref{tab:trials}. We run all simulations first for $t=5\tau$ so that the turbulence can reach a statistically stationary state (usually this takes $t = 2\tau$ but can be up to $t=5\tau$ for strong guide field runs; \citealt{Beattie2022_energybalance,Beattie2021_spdf}) and then average all quantities that we report in this study over the subsequent $t=5\tau$, $5\tau \leq t\leq 10\tau$.
\begin{figure*}[h!]
    \centering
    \includegraphics[width=0.90\linewidth]{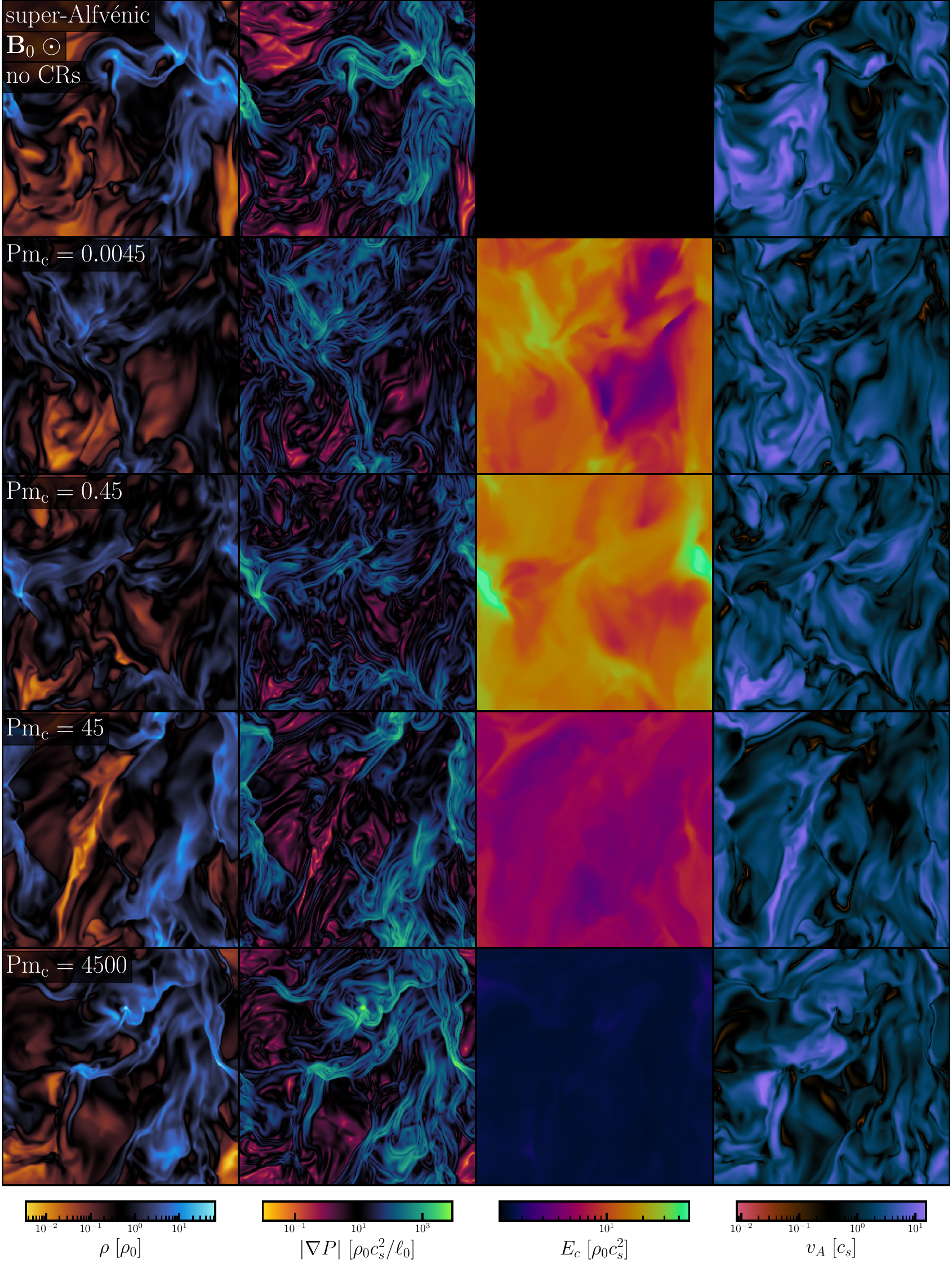}
    \caption{Two-dimensional slice plots through the perpendicular plane to the mean magnetic field, $\bfb_0$, for the full suite of $\mao \approx 10$, super-Alfv\'enic simulations, showing the plasma density, total plasma pressure gradient magnitude, CR energy density, and Alfv\'en velocity magnitude from left to right, respectively. Descending rows show increasing microphysical diffusion compared to the background turbulent diffusion, parameterized by $\Pmcr$. The top row shows the pure MHD simulation (no CR fluid). We note that we keep all colorbars identical between columns, with all the quantities normalized using the thermal units of the plasma. In both the top row and the bottom row the spatial distribution of $E_{\rm{c}}$ is uniform, i.e., $E_{\rm{c}} \approx \rho_0 c_s^2$, which we discuss further in \autoref{sec:heating}.}
    \label{fig:slices}
\end{figure*}

\begin{table*}[t!]
    \centering
    \caption{Simulation parameters}
    \begin{tabular}{cccccccccccc}
    \hline \hline
        run & $\mathcal{M}$ & $\mao$ & $\beta$ & $\rm{Pm_s}$ & $\rm{Pm_{c}}$ & $\rm{Pm}_f$ & $\ell_{v}/L$ &$\ell_{B}/L $ & $\ell_{\rho}/L$& $\ell_{F_{\rm{c}}}/L$  & $N_{\rm grid}^3$  \\ 
        (1) & (2) & (3) & (4) & (5)  & (6) & (7) & (8) & (9) & (10) & (11) & (12) \\[0.2em]
        \hline 
        \texttt{M4MA05noCR}    & 4.3     & 0.72 & 0.056  & - & -  & - & 0.481 & 0.332 & 0.186 & - &  $256^3$      \\ 
        \texttt{M4MA05D22}    & 4.6     & 0.77  & 0.056 & 1.30   & 0.0045  & 0.003  & 0.490 & 0.308 & 0.172& 0.461 & $256^3$     \\ 
       \texttt{M4MA05D24}     & 4.5     & 0.75  & 0.055 & 1.33    & 0.45  & 0.34 & 0.488 & 0.305 & 0.171 & 0.458 & $256^3$      \\ 
        \texttt{M4MA05D26}     & 4.7     & 0.78  & 0.055 & 1.28    & 45  & 35.1 & 0.484 & 0.305& 0.177& 0.442 &$256^3$    \\ 
        \texttt{M4MA05D28}     & 4.8     & 0.80  & 0.055 & 1.25     & 4500  & 3600 & 0.486 & 0.301& 0.181& 0.445 & $256^3$     \\ \hline
        \texttt{M4MA2noCR}    & 3.8     & 2.53 & 0.886 & -  & - & - & 0.506 & 0.302 & 0.214 & - &  $256^3$       \\ 
        \texttt{M4MA2D22}    & 3.7     & 2.47  & 0.891 & 0.40      & 0.0045   &0.011 & 0.503 & 0.295 & 0.205 & 0.383 & $256^3$      \\ 
       \texttt{M4MA2D24}     & 3.7     & 2.47  & 0.891 & 0.40     & 0.45   & 1.11 & 0.507 & 0.307 & 0.209 & 0.395 & $256^3$      \\ 
        \texttt{M4MA2D26}     & 4.1     & 2.73  & 0.886 & 0.36     & 45  & 122.9 & 0.508 & 0.308& 0.214 & 0.326 & $256^3$    \\ 
        \texttt{M4MA2D28}     & 3.6    & 2.40 & 0.888 & 0.42     & 4500   & 10800 & 0.504 & 0.315& 0.209 & 0.314 & $256^3$     \\ \hline
        \texttt{M4MA10noCR}    & 3.3     & 11.0 & 22.22 & -  & -  & - & 0.464 & 0.221 & 0.233 & - &  $256^3$       \\ 
        \texttt{M4MA10D22}    & 3.5     & 11.7  & 22.35 & 0.09    & 0.0045  &  0.05 & 0.495 & 0.224 & 0.238 & 0.395 & $256^3$      \\ 
       \texttt{M4MA10D24}     & 3.3     & 11.0 & 22.22 & 0.09     & 0.45  &  4.95 & 0.483 & 0.216& 0.230 & 0.387 &  $256^3$      \\ 
        \texttt{M4MA10D26}     & 3.5     & 11.7 & 22.35& 0.09     & 45  & 526.5 & 0.476 & 0.210& 0.226& 0.237&  $256^3$    \\ 
        \texttt{M4MA10D28}     & 3.5     & 11.0  & 19.76 & 0.09    & 4500  & 49500 & 0.466& 0.205& 0.205 & 0.226 & $256^3$     \\ \hline
        \hline
    \end{tabular}
    \begin{tablenotes}[para]
        \textit{\textbf{Notes.}} \textbf{Column (1):} the simulation run identity. \textbf{Column (2):} the turbulent Mach number in units of the plasma sound speed, $\mathcal{M}$. \textbf{Column (3):} the mean-field Alfv\'enic Mach number, $\mathcal{M}_{A0}$. \textbf{Column (4):} the isothermal plasma beta $\beta = 2\mao^2/\mathcal{M}^2$ with respect to the mean magnetic field. \textbf{Column (5):} the streaming Prandtl number, $\Pm_{\rm s}$, equal to the $v_{A0} \ell_0$, in units $\sigma_v \ell_0$ (see \autoref{eq:Pms}). \textbf{Column (6):} the diffusion Prandtl number, $\Pmcr$ equal to $\rm{D_{c}}$, the microphysical diffusion coefficient, in units $\sigma_v \ell_0$ (see \autoref{eq:Pms}), \textbf{Column (7):} ratio of streaming to diffusion Prandtl numbers, $\Pm_f = \Pm_{s}/\Pm_{c}$, equal to $v_{A0}\ell_0/\dcr$, describing the contribution between streaming and microphysical diffusion fluxes in \autoref{eqn:flux}. \textbf{Column (8, 9, 10, 11):} the isotropic correlation lengths for the (fluctuating) turbulent field in the velocity $\ell_v$,  magnetic $\ell_B$, density $\ell_\rho$ field, and CR flux, $\ell_{F_{\rm{c}}}$ respectively, in units of $L$ the domain size. See \autoref{eq:corr} and \autoref{sec:corr_scale} for the definition of the correlation scales. \textbf{Column (12):} the number of grid cells used in the discretisation for each simulation.
    \end{tablenotes}
    \vspace{1em}
    \label{tab:trials}
\end{table*}

\subsubsection{Initial conditions and dimensionless quantities}\label{sec:ICs_Pms}
We set an initial homogeneous plasma density field with $\rho_0 = 1$ in code units, and a stationary velocity $|\bfv|=0$. The mean magnetic field $\bfb_0 = B_0 \hat{\bm{z}}$, is oriented only along the $z$ axis. In all of our simulations we initially set $E_{\rm{c}} = 1$ and $\|\bm{F}_{\rm{c}}\| = 0$ in the system units, such that we have an approximate energy equipartition between the CRs and the thermal plasma, which is standard in many CR simulations \citep[e.g.,][]{zweibel2013microphysics, stepanov2014}. 

We define all simulation quantities in terms of the following dimensionless parameters. For the plasma,
\begin{align}
\mathcal{M} &= \sigma_v / c_s, \quad\quad
\mao = \sigma_v / v_{A0},
\end{align}
where $\mathcal{M}$ is the sonic Mach number, $\mao$ is the mean-field Alv\'en mach number, $v_{A0} = B_0 / \sqrt{\rho_0}$ is the mean-field Alfv\'en speed, $B_0$ is the mean magnetic field strength, and $\rho_0$ is the mean plasma density. Note that in our simulation units we have set the magnetic permeability constant, $\mu_0$ to 1, $B_0/\sqrt{\mu_0} \rightarrow B_0$. 

We show the evolution of $\mathcal{M}$ for all runs in \autoref{fig:mach}. The plot shows $\mathcal{M}$ in the final $5\tau$, within the statistically steady state. All runs oscillate around $\mathcal{M} \approx 4$, with some dispersion in $\mathcal{M}$ of order 25\%. This is because we choose to fix the momentum injection into the system (as discussed in \autoref{sec:driving}), across all simulations. The strong mean magnetic field (discussed in more detail in \autoref{sec:ICs_Pms}) changes and redistributes the momentum flux in the turbulence cascade \citep{Zhdankin2017_large_scale_field_e_flux}, hence for a fixed momentum injection, $\mathcal{M}$ changes slightly between the different Alfv\'enic regimes we probe in this study. As in \citet{beattie2020magnetic}, using the steady-state $\mathcal{M}$ and the definition of $\mao$, we set $B_0$ using
\begin{align}
    B_0 = c_s\sqrt{\rho_0} \mathcal{M}\mao^{-1},
\end{align}
allowing us to tune $B_0$ for the exact desired $\mao$. 

The interaction coefficient equation in natural units of the turbulence is
\begin{align} \label{eq:dimensionless_interactio_coeff}
    \hat{\bm{\sigma}}_c^{-1} = \frac{\mathbb{D}_c}{\dcrit} +\  \frac{\ds}{\dcrit}\frac{1}{\sqrt{\hat{\rho}}}\frac{\hat{\bfb} \otimes \hat{\bfb}}{\|\hat{\bfb} \cdot \hat{\bnab} \hat{P_{\rm{c}}}\|}  \cdot(\hat{\mathbb{E}}_c + \hat{\mathbb{P}}_c).
\end{align}
where hatted variables mean that they are scaled by the turbulent correlation lengths, $\sigma_v$, $\rho_0$ or $B_0$ (e.g., $\hat{\bm{\sigma}}_c = \bm{\sigma}_c\ell_0\sigma_v$, $\hat{B} = B/B_0$, etc.; see \autoref{appendix:flux_equation} for the full derivation), $\dcrit$ is as defined in \autoref{eq:turbulent_diffusion} and $\ds = v_{A0} \ell_0$ is an effective diffusion coefficient from the streaming fluxes. This means that there are two dimensionless numbers $\mathbb{D}_c / \dcrit = \dcr/(\sigma_v \ell_0)$ (matching the intuition provided in \citealt{commercon19}) and $\ds/\dcrit = 1/\mao$, each of which defines the importance of the two CR flux contributions. Again we note $\ds/\dcrit = 1/\mao$ is similar to the \emph{critical} $\mao$ provided in \citet{commercon19}, except here we use $v_{A0}$ as the CR propagation speed on the numerator and do not make alterations based on specific ISM phases or ionization states \citep[][see eq 11.]{commercon19}.

Formalizing \autoref{eq:dimensionless_interactio_coeff} further, we may define three dimensionless Prandtl numbers, 
\begin{equation}\label{eq:Pms}
    \Pmcr = \frac{\dcr}{\dcrit}, \quad \Pm_{\rm s} = \frac{\ds}{\dcrit}, \quad \Pm_{\rm f} = \frac{\Pmcr}{\Pm_{\rm s}} = \frac{\dcr}{\ds},
\end{equation}
where $\Pm_{\rm f}$ represents the relative contribution of diffusion, $\Pmcr$, and streaming, $\Pm_{\rm s}$, fluxes to the magnitude of $\bm{\hat \sigma}_c^{-1}$. For $\Pm_{\rm f} > 1$ the diffusive fluxes dominate $\bm{\hat \sigma}_c^{-1}$, and for $\Pm_{\rm f} < 1$, the streaming fluxes dominate. We report the values of $\Pm_{\rm f}$ for all simulations in \autoref{tab:trials}. Because both $\Pmcr$ and $\Pm_{\rm s}$ compare either the CR diffusion or the streaming-related diffusion with the turbulence diffusion, both $\Pmcr$ and $\Pm_{\rm s}$ parameterize how strongly coupled $\bm{\hat \sigma}_c^{-1}$ is in the flux and plasma momentum equation. We will discuss this in much more detail throughout this paper.

We run a suite of simulations varying $\Pm_{\rm s}$ (the streaming flux contribution to $\bm{\hat \sigma}_c^{-1}$), and $\Pmcr$ (the diffusion flux contribution) to explore the backreaction (or coupling) effect on the turbulent plasma. We run simulations across three Alfv\'enic Mach number regimes, sub- ($\mao \approx 0.75$; streaming dominant compared to the turbulence), trans- ($\mao \approx 2.5$; streaming and the turbulence are comparable), and super-Alfv\'enic ($\mao \approx 11$; the turbulence dominates over the streaming), with the exact values for $\mao$ for each simulation shown in \autoref{tab:trials}. For each $\mao$ configuration, we run 5 sets of simulations, where we vary $\Pm_{\rm c} =D_{\rm c,\parallel}/\dcrit \in 4.5 \times \{0, 10^{-3}, 10^{-1}, 10^{1}, 10^{3}\}$ with a $D_{\rm c,\parallel}/D_{\rm c,\perp}$ ratio of $10^{-4}$ in all runs. We report all values of $\Pm_{\rm c}$ in \autoref{tab:trials}.

\begin{figure*}[hbt!]
    \centering
    \includegraphics[width=0.94\textwidth]{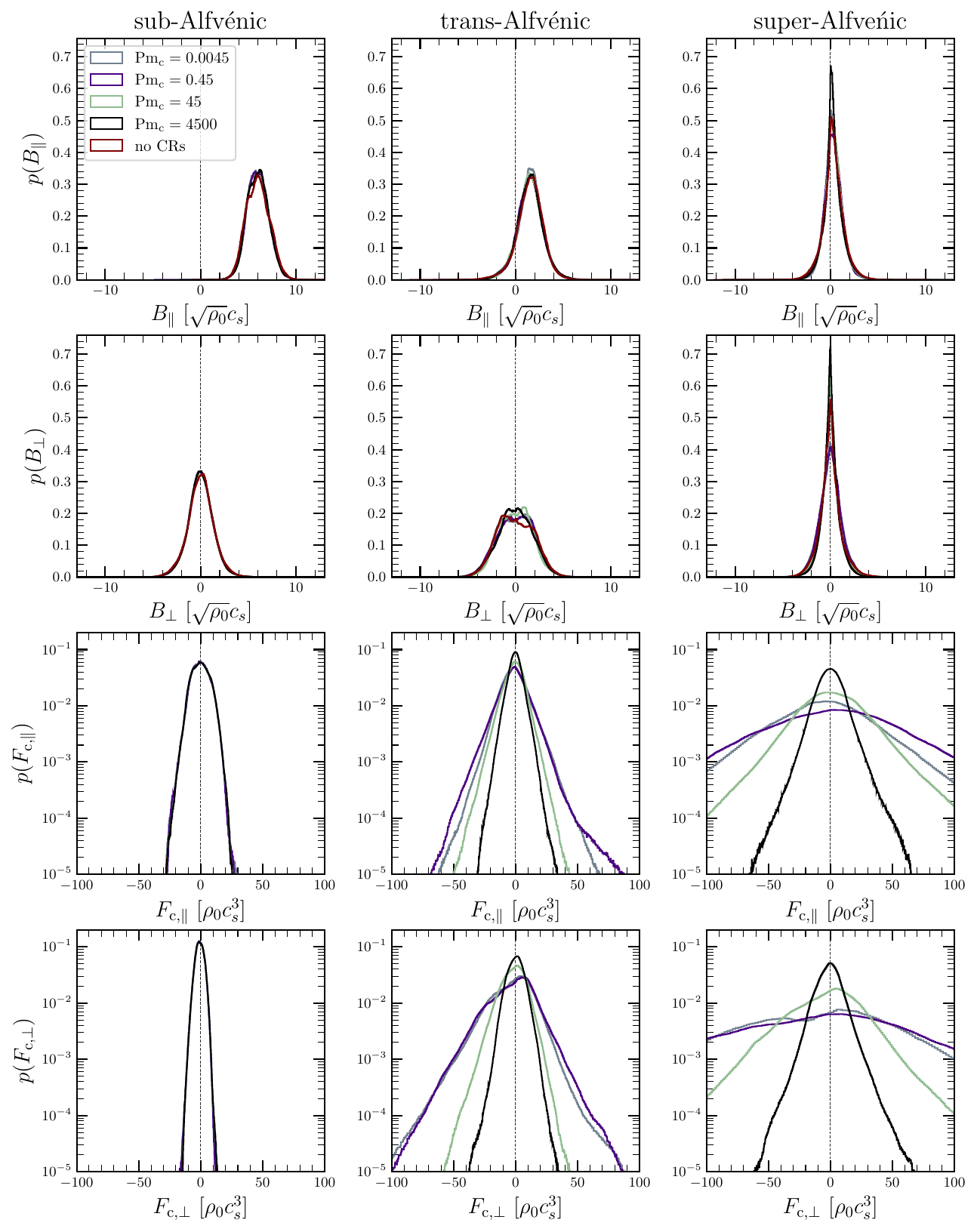}
    \caption{\textbf{Top row}: One-dimensional probability density distributions of the magnetic field strength in the direction parallel to the mean magnetic field, $\bm{B}_{\parallel}$. \textbf{Second row:} the same but for the perpendicular component,$\bm{B}_{\perp}$). \textbf{Third row}: the parallel component of the CR flux, $\bm{F}_{\parallel}$. \textbf{Fourth row:} the perpendicular component of the CR flux, $\bm{F}_{\perp}$. All variables have been non-dimensionalized by the thermal velocities, $c_s$ and mean plasma densities, $\rho_0$. The columns show sub, trans, and super-Aflv\'enic runs, from left to right respectively. The color indicates the value of the microphysical CR diffusion coefficient compared to the turbulent diffusion, $\Pmcr$ (see \autoref{eq:Pms}), with a pure MHD simulation (no CRs) being run and shown in red.}
    \label{fig:1Dpdf}
\end{figure*}

\subsection{Field visualizations}
\autoref{fig:slices} shows a series of two-dimensional slice plots from our suite of $\mao \approx 10$, super-Alfv\'enic simulations at $t=6\tau$ (see plot \autoref{fig:mach}). We show the plasma density, total pressure gradient magnitude of $P=P_{\rm thermal} + P_{\rm mag}$, CR energy density, and Alfv\'en speed with the rows representing increasing levels of $\Pmcr$ (i.e., microphysical diffusion compared to the background turbulence diffusion). We note as there is no CR fluid in the simulations in the top row, hence the third panel is set to an arbitrary constant. 

Very little variation is seen between the simulations in \autoref{fig:slices}, in any of the fields, with the exception of $E_{\rm{c}}$ in the third column. As one may expect, when $\Pmcr \gg 1$, the $E_{\rm{c}}$ field is uniform, since $|\nabla E_{\rm{c}}|$ is quickly smoothed out by the diffusion happening on shorter timescales than the turbulence, similar to what \citet{commercon19} showed. However, note that there are some differences between this plot and the results shown in Figure~3 of \citet{commercon19}, who find a clear variation in plasma density with increasing $\Pmcr$, noting that in physical units, the microphysical diffusion coefficients are in the same range as in \citet{commercon19}. We explore these differences further in \autoref{sec:pdfs} and \autoref{sec:heating}. 

\section{Probability density functions}
\label{sec:pdfs}

We perform experiments similar to \citet{commercon19}, statistically comparing field quantities in our plasma for different values of $\Pmcr$. We note that in the \citet{commercon19} transport model there is no CR streaming, whereas in this study, for all simulations that include a CR fluid, we include streaming as defined in \autoref{eq:interaction} in our transport model, introducing a new dimensionless parameter, $\Pm_{\rm s}$. First we explore the simplest statistics -- the one dimensional distribution functions of the CRMHD turbulence.

\subsection{One-dimensional PDFs}

\begin{figure*}[hbt!]
    \centering
    \includegraphics[width=0.98\textwidth]{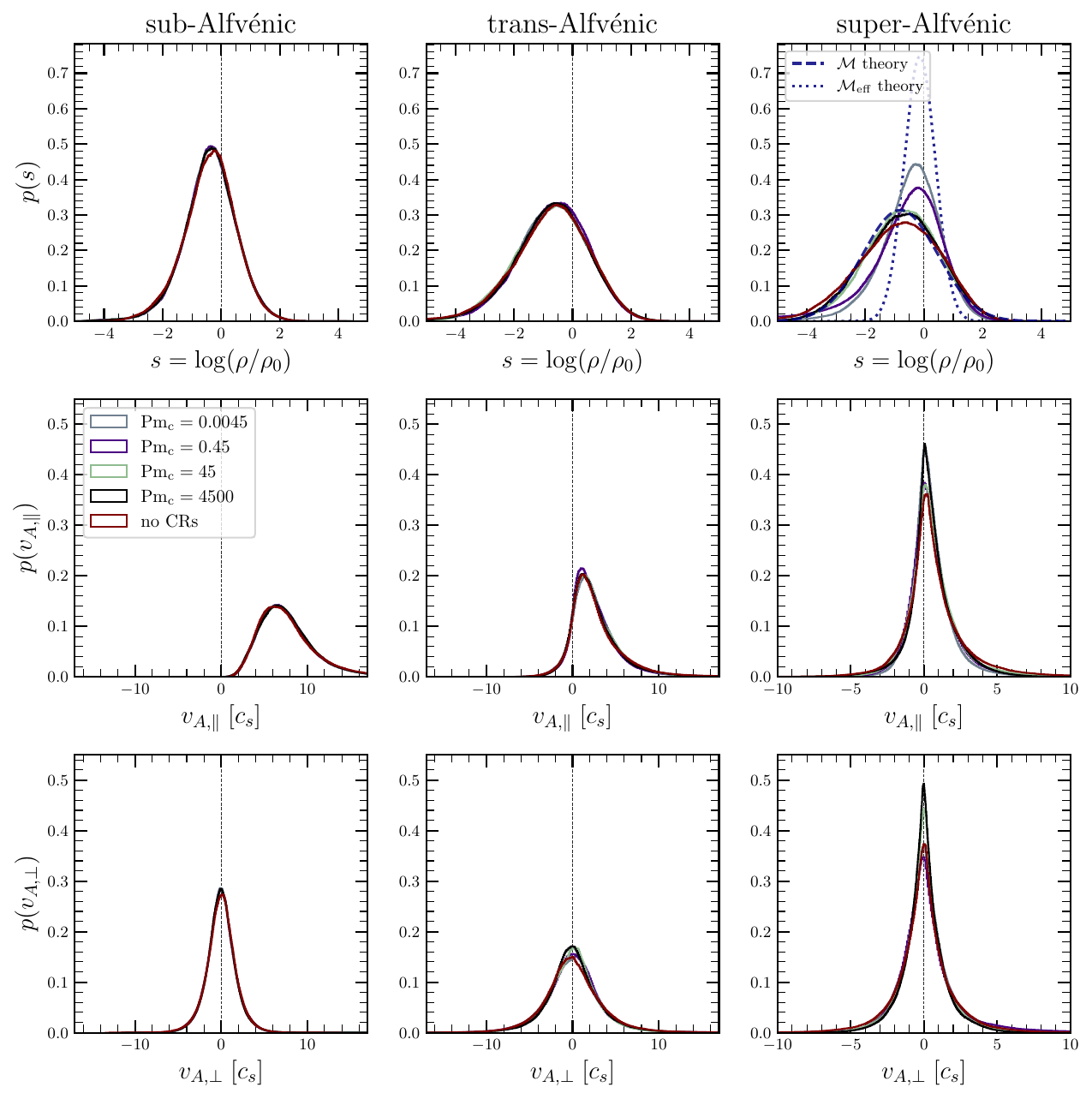}
    \caption{\textbf{Top row:} The same as \autoref{fig:1Dpdf} for  the logarithmic, mean-normalized plasma density, $s=\log(\rho/\rho_0)$. In the top-right panel, we show lognormal models for the density fluctuations for mixed-driving $\mathcal{M}\approx 4$ turbulence without any influence from the CRs ($\mathcal{M}$ theory) and with an effect from the CRs by changing the plasma effective sound speed directly from the CR $P_{\rm{c}} \propto \rho^{4/3}$ equation of state ($\mathcal{M}_{\rm eff}$ theory), showing that the density fluctuations are bounded between these limits. \textbf{Second row:} The parallel components of the Alfv\'en velocity, $\bm{v}_{A,\parallel} = \bm{B}_{\parallel} / \sqrt{\rho}$. \textbf{Bottom row:} The perpendicular components of the Alfv\'en velocity, $\bm{v}_{A,\perp} = \bm{B}_{\perp} / \sqrt{\rho}$. \textbf{Note:} the limits for the x-axis of the rightmost $v_A$ panels have been reduced for better visual comparison between runs in the super-Alfv\'enic regime. }
    \label{fig:2pdf}
\end{figure*}

In \autoref{fig:1Dpdf} we show one-dimensional probability density distribution functions (PDFs) of the component of the $\bfb$-field in directions parallel ($B_z = B_\parallel$) and perpendicular ($B_x = B_\perp$) to the guide field $\bm{B}_0 = B_0 \hat{\bm{z}}$, as well as the CR flux ($\bff_{\rm{c}}$) for the same components. 

We see that for all our runs, there is very minimal noticeable effect of the CR fluid, irrespective of $\Pmcr$, on the distribution of the $\bfb$-field along either the parallel or perpendicular directions with respect to the mean field. For our sub-Alfv\'enic runs, we also have no variation of the CR flux distribution along either component as seen in the left-most bottom 2 rows of \autoref{fig:1Dpdf} as we alter $\Pmcr$. We do see in both the sub and trans-Alfv\'enic runs an anisotropy between the CR flux components both parallel and perpendicular to the guide field which is expected due to in part the stronger streaming ($\Pm_{\rm s} \sim 1.3$ - for sub-Alfv\'enic, and $\Pm_{\rm s} \sim 0.4$ for trans-Alfv\'enic) in these regimes, and also from the stronger, and more tightly aligned (to $z$) guide field.

The component-wise flux distributions in the trans and super-Alfv\'enic runs of \autoref{fig:1Dpdf} also show a clear effect of $\Pmcr$ with increasing values initially increasing the width of the PDFs in both direction. However, in the strong diffusion limit ($\Pmcr=4500$ ), we see a narrowing of the PDFs for both these regimes. 

In the weak mean field case (super-Alfv\'enic), shown in the right-most column, we see very minor difference in the PDFs for the components of the CR flux between the parallel and perpendicular directions. This is to be expected, as for super-Alfv\'enic turbulence, the field becomes globally isotropic \citep{Beattie2022_energybalance}, though locally anisotropic, even when $B/B_0 \gg 1$ \citep{Fielding2023_plasmoids}.

\autoref{fig:2pdf} shows the one-dimensional PDFs for the mean-normalized, logarithmic plasma density, $s\equiv\log(\rho/\rho_0)$, which we expect to be approximately Gaussian \citep{Federrath2008,Hopkins2013_spdf,Squire2017_s_pdf,Mocz2019_markov_model,Beattie2021_spdf}, and the parallel and perpendicular components of the Alfv\'en speed in $c_s$ units. We see from the upper row of \autoref{fig:2pdf} that for $\mao  \lesssim 2.5$, i.e., trans-to-sub-Alfv\'enic, where $\Pm_{\rm s} \sim 1$, $\Pmcr$ (even in the most extreme case where there is no CR fluid) has no effect on the $s$ distribution, which is different than \citet{commercon19}, who found that reducing the value of $\dcr$ (our $\Pmcr$) causes a reduction in the width of these PDFs. However, in the super-Alfv\'enic, $\mao  \gtrsim 2.5$ regime, there is a clear inverse relation between the width of the $s$ PDF (top right panel) and $\Pmcr$, which we explain further below.

It is known that the variance of the $s$ PDF is $\sigma_s^2 = \log(1 + b^2\mathcal{M}^2)$ in hydrodynamical turbulence, and $\sigma_s^2 = \log(1 + b^2\mathcal{M}^2(\beta / (\beta+1)))$, where $\beta = P_{\rm therm}/P_{\rm mag}$ is the plasma beta of the plasma and $b$ is the driving parameter, in super-Alfv\'enic mean field, magnetohydrodynamic turbulence  \citep{Federrath2008,Federrath2009,Federrath2010,Molina2012_dens_var,jin2017effective,beattie2021multishock,Beattie2021_spdf}. If the plasma and the CR fluid are strongly coupled this modifies the equation of state of the plasma \citep{salem2014cosmological,commercon19}, which will change $\sigma_s^2$. We will directly measure the equation of state in the following section. 

For a relativistic fluid, like the CRs, the equation of state is $P_{\rm{c}} \propto \rho^{4/3}$. Hence, when the plasma inherits such an equation of state, as it does when $\Pmcr$ is low \citep{commercon19}, the effective sound speed and $\mathcal{M}$ become,
\begin{align}
    c_{\rm eff} = \sqrt{ \frac{4}{3}\frac{P_{\rm{c}}}{\rho} + \gamma \frac{P_{\rm therm}}{\rho}}, \quad\quad \mathcal{M}_{\rm eff} = \frac{\sigma_v}{c_{\rm eff}},
\end{align}
for our isothermal plasma with $c_s = 1 \iff \gamma P_{\rm therm}/\rho = 1$. Because $c_{\rm eff} > c_s$, $\mathcal{M}_{\rm eff} < \mathcal{M}$, in turn reducing $\sigma_s^2$, i.e., the plasma becomes more homogeneous with less variation between over/under dense regions through the reduction of its $\mathcal{M}_{\rm eff}$ number (increased incompressibility) \citep{Federrath2008,Molina2012_dens_var,Robertson2018_dense_regions}. In the top-right panel of \autoref{fig:2pdf}, we plot two theoretical lognormal models for the $s$ PDF in \autoref{fig:2pdf}. The first, using
\begin{align}
    \sigma_s^2 = \log\left(1 + b^2\mathcal{M}^2\frac{\beta}{\beta+1}\right), \quad\quad\text{($\mathcal{M}$ theory)}
\end{align}
and the second 
\begin{align}
    \sigma_s^2 = \log\left(1 + b^2\mathcal{M}_{\rm eff}^2\frac{\beta}{\beta+1}\right), \quad\quad \text{($\mathcal{M}_{\rm eff}$ theory)}
\end{align}
where we use $b = 0.5$ to match the $\chi$ for our driving \citep{Federrath2010} and $\beta = 2 \mao^2/\mathcal{M}^2$ for an isothermal EOS\footnote{Note that this comparison to the isothermal theories is only meant to define an upper and lower bound for which the data may be situated in. For example, we know that there will be modifications to even the basic $\sigma _s^2 \propto \mathcal{M}^2$ relations for an adiabatic EOS \citep[see Equation~10 in ][]{Nolan2015_density_variance}.}. Further, we utilize the fact that $\left\langle s \right\rangle = - \sigma_s^2/2$ for a lognormal model, where $\left\langle s \right\rangle$ is the mean log density (i.e., the lognormal model is a one-parameter model, solely a function of $\sigma_s^2$). Both PDFs bound the data well, with the $\mathcal{M}_{\rm eff}$ model setting a lower bound for the variance  of the PDF, and the hydrodynamical $\mathcal{M}$ model setting an upper bound.

As we note above, for $\Pm_{\rm s} < 1$, a weakly diffuse ($\Pmcr < 1$) CR fluid acts to smooth out the plasma density fluctuations through its faster $c_{\rm eff} > c_s$ acoustic response. As CR streaming ($\Pm_{\rm s} > 1$) alone is strong enough to decoupling the CR fluid from the plasma when $\mao \lesssim 2$, we do not see any effect of the CR fluid on the plasma even for very low $\Pmcr$ values. This can be understood from the interaction term described in \autoref{eq:interaction}, which couples the CR fluid and plasma through the momentum equation in \autoref{eq:CRMHD-mom}. Strong streaming, $\Pm_{\rm s} > 1$, or strong levels of diffusion $\Pmcr > 1$ act to reduce the strength of this coupling $\bm{\sigma}_c \sim 1/\Pmcr$ or $\bm{\sigma}_c \sim 1/\Pm_{\rm s}$, depending upon $\Pm_{\rm f}$. We explore this in more detail, in the following sections.

The $v_A$ PDFs shown in the second and third rows of \autoref{fig:2pdf} are all mostly unaffected by $\Pmcr$ with the exception of the super-Alfv\'enic runs (right-most columns). This is expected for the sub- and trans-Alfv\'enic regimes where both the $\bfb$-component distribution, and the $s$ distributions are unaffected by $\Pmcr$. Hence from the definition of $\bfv_A$ its distribution should also be unaffected. We do see dependence of the $s$ distribution on $\Pmcr$ in the super-Alfv\'enic runs, and hence because $\bfv_A \propto 1/\sqrt{\rho}$ we do see that there is some variation in $\bfv_A$ as a function of $\Pmcr$ in this regime.

\subsection{Two-dimensional PDFs}
\begin{figure*}
    \centering
    \includegraphics[width=0.98\textwidth]{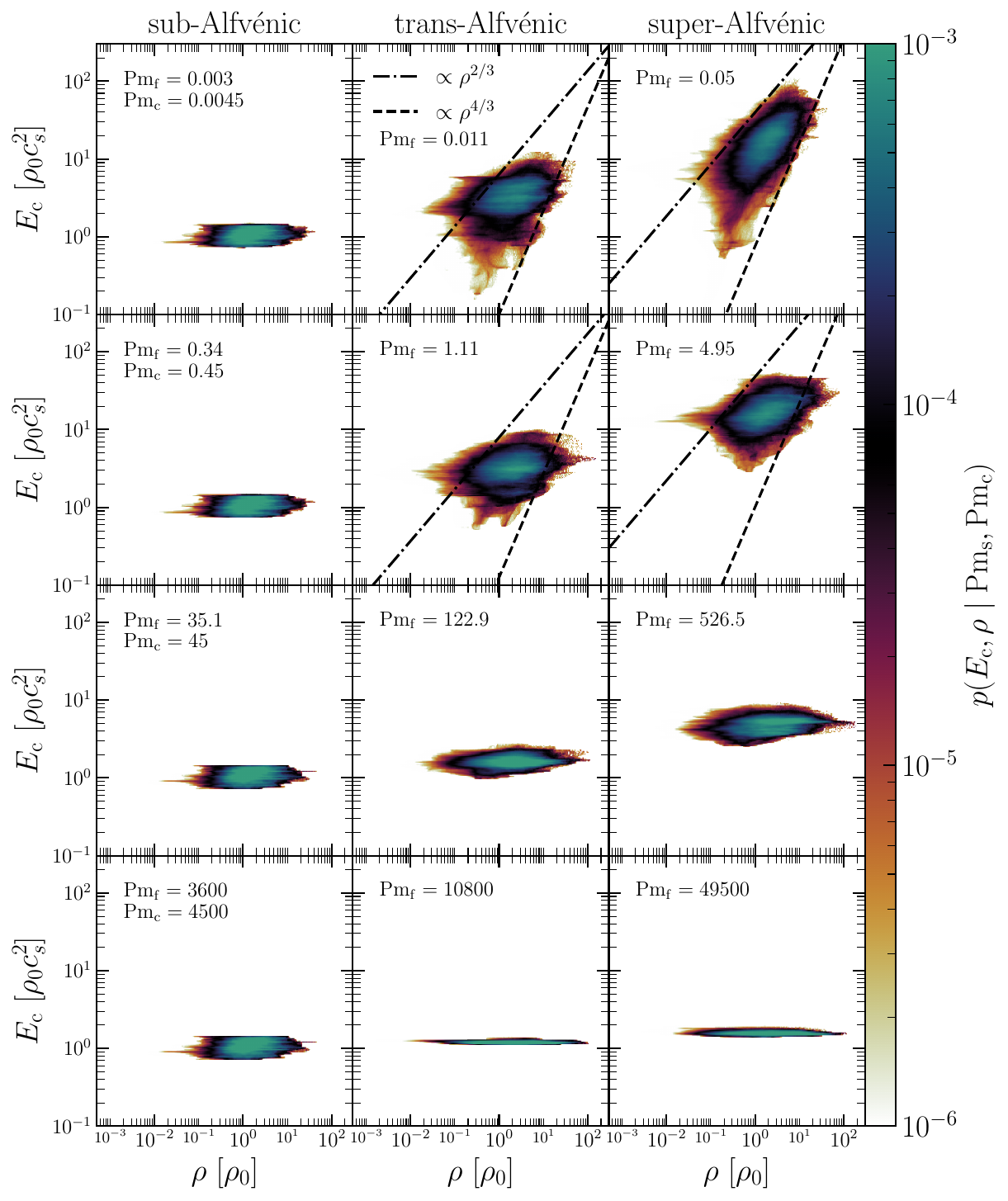}
    \caption{Two-dimensional probability density functions of cosmic ray energy density, $E_{\rm{c}}$, and plasma density, $\rho$, $p(E_{\rm{c}}, \rho | \Pm_{\rm s}, \Pmcr)$, with the probability density shown via the color, and conditional over specific choices of $ \Pm_{\rm s}$ and $\Pmcr$. The rows show runs with increasing microphysical diffusion Prandtl numbers $\Pmcr \in \{0.0045, 0.45, 45, 4500 \}\;$ from top to bottom. The columns show increasing $\mao$ for sub-, trans-, to super-Alfv\'enic turbulence from left to right, respectively. In the top two rows, and two right-most columns (strong coupling regime) we plot $E_{\rm{c}} \propto \rho^{4/3}$, and $E_{\rm{c}} \propto \rho^{2/3}$ trendlines, representing the equation of state of a relativistic fluid, and the modifications to the equation of state through streaming, respectively. }
    \label{fig:2d_pdf_energy_density}
\end{figure*}

\begin{figure*}
    \centering
    \includegraphics[width=1\textwidth]{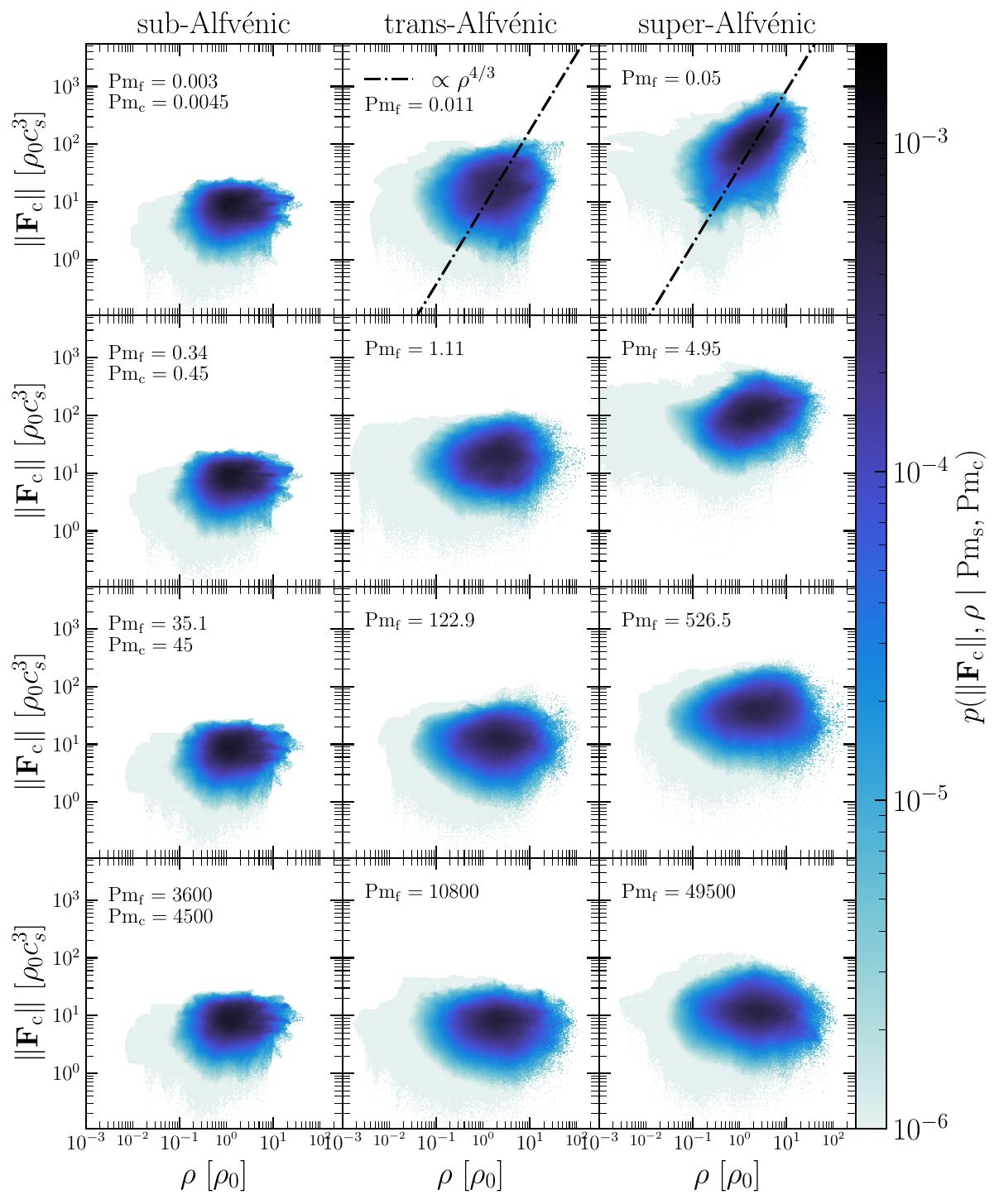}
    \caption{The same as \autoref{fig:2d_pdf_energy_density} except for the two-dimensional probability density function of the CR flux magnitude $\| \bm{F}_{\rm{c}} \|$ and $\rho$. In the strong coupling regime (top-right panels), we observe $\| \bm{F}_{\rm{c}} \| \propto \rho^{4/3}$ (see \autoref{eq:derive_flux_dens}).}
    \label{fig:2d_pdf_flux_density}
\end{figure*}

We plot the two-dimensional distribution functions of $E_{c}$ and $\|\bff_{\rm{c}}\|$ and the plasma density in \autoref{fig:2d_pdf_energy_density} and \autoref{fig:2d_pdf_flux_density}, respectively. The descending rows show increasing values of $\Pmcr$, and the columns showing increasing $\mao$ (or decreasing $\Pm_{\rm s}$) from left to right. In both \autoref{fig:2d_pdf_energy_density} and \autoref{fig:2d_pdf_flux_density}, the color is indicative of the volume-weighted probability density (e.g., darker probability densities correspond to the most volume-filling state in the CRMHD plasma). In the weak diffusion $\Pmcr < 1$, weak streaming $\Pm_{\rm s} < 1$ simulations, we plot two lines in black showing the analytical relations between $E_{\rm{c}} \sim P_{\rm{c}}$ and $\rho$ coming directly from the adiabatic index of the relativistic CR fluid without diffusion $P_{\rm{c}} \sim E_{\rm{c}} \propto \rho^{4/3}$ and correction to the equation of state based on the streaming fluxes $E_{\rm{c}} \propto \rho^{2/3}$ \citep{wiener2017b, kempski20, quataert22}.

We see in the first column of \autoref{fig:2d_pdf_energy_density}, in the sub-Alfv\'enic regime, there appears to be no correlation between the $\rho$ and $E_{\rm{c}}$, even for very small values of $\Pmcr$, i.e., low values of microphysical CR diffusion compared to the turbulent diffusion, $\Pmcr < 1 \iff \dcr < \dcrit  $. In this regime, the plasma and CR fluid are effectively decoupled, but not via the diffusive fluxes, which we explain quantitatively in more detail throughout the main results of this study. In \cite{Dubois2019}, it was also found that CR streaming produces a more uniform $E_{\rm{c}}$ distribution, as compared to no streaming runs. This reduces the magnitude of $\nabla P_{\rm{c}}$ effectively acting to decouple the fluid as seen here, and explained in more detail in \autoref{sec:heating}. 

In contrast, the top-right panel, representing the super-Alfv\'enic, low-diffusion simulation, where $\Pmcr < 1$ and $\Pm_{\rm s} < 1$, we observe strong coupling. The low- and high-density bounds of the two-dimensional PDF are well described by the theoretical relations $E_{\rm{c}} \propto \rho^{2/3}$ and $E_{\rm{c}} \propto \rho^{4/3}$, respectively. This implies that (1) strong CR-plasma coupling exists in this regime and (2) the nature of the coupling is due to a mixture of both streaming and diffusion changing the $P_{\rm{c}}$ and hence, $E_{\rm{c}}$. The level of agreement between the measured PDFs and the analytical limits, hence the strength of the CR-plasma coupling, becomes worse / weaker as either the $\Pmcr$ increases, and/or as $\Pm_{\rm s}$ increases because both large parameter limits independently act to decouple the CR fluid. As $\Pm_{\rm s}$ increases, $v_{A0}$ increases and we transition from super to sub-Alfv\'enic turbulence with $\Pm_{\rm s} > 1$ which alone will decouple the fluids. At $\Pmcr \approx 45$, we find a completely uniform distribution across $\rho$ even for $\Pm_{\rm s} < 1$ plasmas because, as for $\Pmcr > 1$, the diffusion time is shorter than the turbulence turnover time of the plasma, decoupling the CR fluid from the plasma. 

We also note a difference in average $E_{\rm{c}}$ across the regimes, where we see in the more coupled (weak diffusion weak $\bfb$-field) runs that there is a higher $\langle E_{\rm{c}} \rangle$ with the \texttt{M4MA10D24} simulation having the highest average CR energy. This is further explored in \autoref{sec:heating}.

In \autoref{fig:2d_pdf_flux_density} we plot the two-distribution functions of  $\|\bff_{\rm{c}}\|$ and $\rho$, similarly as in \autoref{fig:2d_pdf_energy_density}. In general, the distribution of $\|\bff_{\rm{c}}\|$ is much less correlated with $\rho$ in all regimes, compared to $E_{\rm{c}}$. However, we find a similar relation between $\|\bff_{\rm{c}}\|$ and gas density with $\|\bff_{\rm{c}}\| \propto \rho^{4/3}$ in the weak-diffusion, weak-field (top right) limit. This is expected in the turbulent diffusion-dominated limit of the steady-state CR flux because, 
\begin{align}\label{eq:derive_flux_dens}
    F_{\rm{c}} \sim P_{\rm{c}}v \sim \rho^{4/3}v.
\end{align}
In the sub-Alfv\'enic ($\mao < 1)$ regime we see almost no variation in the distributions across $\Pmcr$ values, with the only significant variations in the weak $\bfb$-field, weak diffusion limit. This is due to the large fluxes, with $\Pm_{\rm s}\sim 1.3 \equiv \mao \sim 0.75$ for all runs in this regime. From \autoref{eq:dimensionless_interactio_coeff} we can see that when $\Pm_{\rm f} < 1$ $\bm{\sigma}_c \sim 1/\Pm_{\rm s}$ and the coupling strength is decreased between the CR fluid and the plasma, purely by the strong streaming fluxes. As we show in \autoref{tab:trials}, for the sub-Alfv\'enic regime, $\Pm_{\rm s} > 1$ when $\Pm_{\rm f} < 1$, which means that the CR and plasma will always tend towards a decoupled state. Even when $\Pm_{\rm f} > 1$, and the diffusive fluxes begin to become strong compared to the streaming, $\Pmcr>1$, which then decouples the CR fluid and the plasma via diffusion. Hence, nowhere in the sub-Alfv\'enic turbulence do the CRs and plasma couple.  

\section{Power spectrum and correlation scales}
\label{sec:corr_scale}
To probe how the different CRMHD fluid regimes change the nature of the plasma turbulence we explore the power spectra and correlation (or outer) scales of the plasma density, velocity field, magnetic field, and the CR flux.

\begin{figure*}
    \centering
    \includegraphics[width=0.97\linewidth]{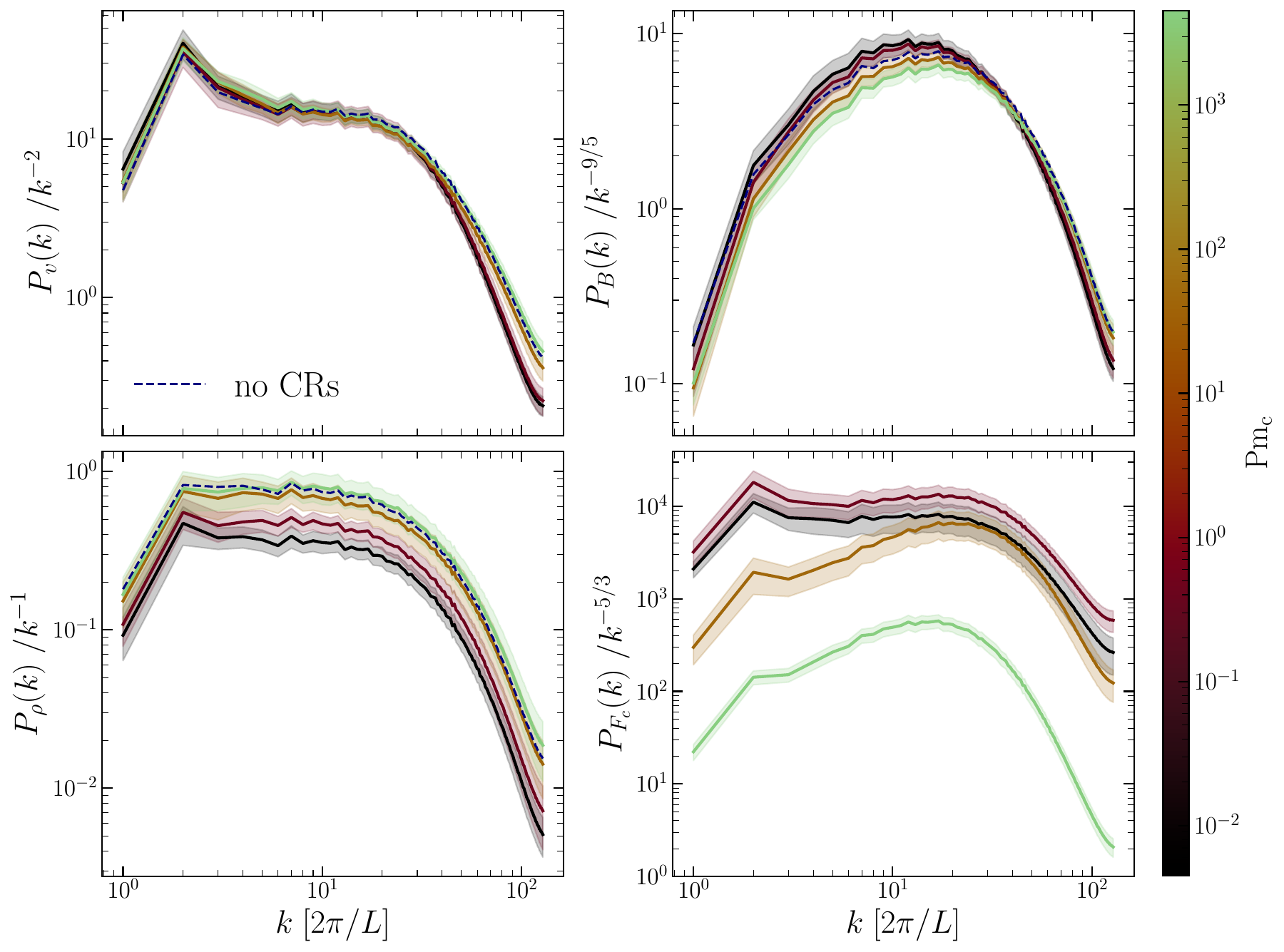}
    \caption{Isotropic power spectrum for the super-Alfv\'enic runs for $\bfv$, $\bfb$, $\rho$, and $\bff_{\rm{c}}$ with the field indicated in the subscript of the $P(k)$ label. We average over $5\tau$ with the mean in solid and the standard deviation indicated in the shading. We compensate the velocity spectra by $k^{-2}$ for easier comparison to $P_v(k) \propto k^{-2}$ \citet{Burgers1948_turbulence} turbulence, found on the supersonic wavemodes within supersonic turbulence \citep{Federrath2021_sonic_scale,Cernetic2024_supersonic_turbulence_GPU,Beattie2025_nature_astronomy}. With the color, we show the range of $\Pmcr = \dcr/\dcrit$, varying from weak microphysical diffusion $\Pmcr \ll 1$, and hence strong CR-plasma coupling (see \autoref{fig:2d_pdf_energy_density}), to strong microphysical diffusion $\Pmcr \gg 1$ and hence weak CR-plasma coupling. We also show a no CR (pure MHD) simulation in the blue dashed line. Overall, the turbulence in the velocity and magnetic fields remain approximately invariant, and the CR-plasma coupling mostly impacts the plasma density and CR fluid flux fluctuations.}
    \label{fig:power-spectra}
\end{figure*}

\subsection{Definitions}
\label{sec:definition_corr}
The correlation scale describes the largest scale over which two-point correlations exist (e.g., the outer-scale or integral scale of the turbulence in a regular turbulent cascade). As in the textbook definition, we define the correlation length $\ell_{\bm{X}}$ as
\begin{equation}
\label{eq:corr}
    \frac{\ell_{X}}{L} = \frac{\displaystyle\int k^{-1}P_{X}(k)\; dk }{\displaystyle\int P_{X}(k) \;dk},
\end{equation}
where $P_{\bm{X}}(\bm{k}) \equiv \|\tilde{\bm{X}}(\bm{k})\tilde{\bm{X}}^{\dagger}(\bm{k})\|$ is the power spectrum for field $\bm{X}$, where $\tilde{\bm{X}}$ is the Fourier transform of vector field $\bm{X}$ and $\tilde{\bm{X}}^{\dagger}$ is the complex conjugate transpose. The isotropic correlation lengths (which we report) are computed by integrating the initial 3D power spectrum $P(\bm{k})$ over spherical shells
\begin{align} \label{eq:isotropic_power}
    P(k) = \int_{\Omega_k} P(\bm{k}) \; k^2 d\Omega_k
\end{align}
where $\Omega_k$ is the solid angle at fixed $k = |\bm{k}|$ and then $P(k)$ is substituted into \autoref{eq:corr}. For all fields, we ensure that there is no $k_i = 0$ mode by subtracting the mean in real-space, which we find can leak into other low-$k$ modes (see e.g., \citealt{bustard23}, where the mean-field $k=0$ leaks into a broad band of low-$k$ modes in their magnetic spectrum), significantly reducing $\ell_X$.

\subsection{Power spectra}
In \autoref{fig:power-spectra} we plot the four isotropic power spectra for the $\mao \approx 10$ runs, with the color indicating the value of $\Pmcr$, probing the turbulence across the weak-to-strong coupling regime between the CR fluid and the plasma. Firstly, as we noted in \autoref{sec:methods}, these simulations are significantly limited in Reynolds numbers, with $\rm{Re} \approx 500-1000$, hence we do not expect an extended cascade, or a fully-developed cascade at all (e.g., a self-similar range of $k$ completely independent of driving and numerical dissipation). For $\M \approx 4$ on the outer-scale, the transition to subsonic turbulence happens at $k_s L/2\pi \approx 30$ \citep{Beattie2025_nature_astronomy}, which, as evidenced in \autoref{fig:power-spectra}, is well within the dissipation range for these simulations. Hence, we only properly resolve a handful of large-scale, supersonic modes. However, we find that all of the spectra, except for $P_B(k)$, become self-similar $P(k) \propto k^\beta$.

$P_v(k)$ begins to flatten under the $k^{-2}$ compensation, which is expected for a spectrum dominated by singularities, such as \citet{Burgers1948_turbulence}-type turbulence. On the low-$k$ modes, we see very minor differences between the $P_v(k)$ for different $\Pmcr$, i.e., different CR-plasma coupling strengths, indicating at first glance that the velocity structure is not being changed by the CR fluid coupling. Although the $P_v(k)/k^{-2} = \rm{const.}$ range is extremely limited, this suggests that a strongly coupled CR fluid does not change the velocity-dispersion length-scale relation, which is $\sigma_v \propto \ell^{1/2}$ for a $P_v(k) \propto k^{-2}$ spectrum \citep{Larson1981_LarsonsLaw}. The strongest coupling effects are in the high-$k$ modes of $P_v(k)$, where the most viscous $k$ are being decayed with a steeper exponential, but this may be strongly influenced by numerical effects because there is no explicit viscosity. We explore this further through a Hodge-Helmholtz decomposition to separate the compressible and solenoidal modes in \autoref{sec:heating}.
We compensate $P_B(k)$ by $k^{-9/5}$, an empirical relation found in extremely high-resolution MHD turbulence simulations \citep{Fielding2023_plasmoids,Beattie2025_nature_astronomy}. We find that under this compensation $P_B(k)$ flattens, but many high-resolution simulations have shown that $P_B(k)$ does not become completely self-similar until $\rm{Re} \gtrsim 10^4$ \citep{Fielding2023_plasmoids}, and the Alfv\'en scale (energy equipartition scale), which defines the outer-scale of the magnetic cascade, is well-resolved, \citep{Beattie2025_nature_astronomy}. For $\mathcal{M}\approx 4$, this means one needs $N_{\rm grid} \gtrsim 2,\!000$, and lack of a power law in the spectrum is therefore expected at $N_{\rm grid} = 256$. Similarly to $P_v(k)$, we see no significant variation in $P_B(k)$ as we change $\Pmcr$, hence both the magnetic field and the velocity field (the classical MHD turbulence variables; \citealt{Schekochihin2022_a_bias_review}) are rather invariant to the coupling strength of the CR fluid. Now we turn to the fields that are a function of $\Pmcr$.

Both $P_\rho(k)$, where we have taken the power spectrum of $\rho/\rho_0 - 1$, and $P_{F_{\rm{c}}}(k)$ vary as a function of $\Pmcr$. $P_\rho(k)$ (bottom-left panel) varies only in integral, noting that from Parseval's theorem,
\begin{align}\label{eq:spectra_integral}
    \left\langle (\rho/\rho_0)^2 \right\rangle = \int^\infty_0 P_\rho(k) dk.
\end{align}
We find that $P_\rho(k)$ monotonically decreases with decreasing $\Pmcr$, as the CR-plasma becomes more coupled. As we show in the $\log(\rho/\rho_0)$ PDFs in \autoref{fig:1Dpdf}, the integral of $P_\rho(k)$ has to decrease, but what we find is that it decreases equally across all $k$ modes as the coupling becomes stronger and all $P_\rho(k)$, regardless of $\Pmcr$, follow a $P_\rho(k) \propto k^{-1}$ spectrum. Such shallow $P_\rho(k)$ have been previously found in supersonic turbulence \citep{Kim2005_density_spectrum}. This means that all $\rho$ fluctuations decrease in magnitude as the CRs change the equation of state of the plasma, but the correlation scale remains invariant, which is subtly different from changing $\mathcal{M}$ in the forcing function, which changes both $\left\langle (\rho/\rho_0)^2 \right\rangle$ and the correlation scale of the mass density \citep{Kim2005_density_spectrum,Beattie_2020_filaments_and_striations,Beattie2022_va_fluctuations}. Although we probe different plasma regimes than \citet{bustard23}, it is still interesting to compare our findings regarding the dependence of the spectrum on $\Pmcr$. In \citet{bustard23} the kinetic energy spectra $P_{\sqrt{\rho} v}(k)$ are measured, with differences being found in the spectrum for different levels of microphysical diffusion. Our results suggest these differences in $P_{E_k}(k)$ may be predominately due to changes to $P_\rho(k)$ with the structure, and magnitude of the pure velocity field being mostly unaffected. We do note however that there may be differencesin subsonic turbulence (as in \citealt{bustard23}) compared to supersonic turbulence (this work), as well as the different turbulent forcing mechanisms as in \citet{bustard23} the forcing acts purely on the compressive modes.

$P_{F_{\rm{c}}}(k)$, shown in the bottom-right panel, compensated by $k^{-5/3}$ to guide the eye, varies both in integral (as in \autoref{eq:spectra_integral}) and in structure (correlation scale) for different values of $\Pmcr$. Qualitatively, the flux fluctuations vary roughly between the structure in $P_B(k)$ in the weakly-coupled $\Pmcr \gg 1$ limit (the CR fluid traces the magnetic field lines via \autoref{eq:diffusion_coefficient}), and then in $P_v(k)$ in the strongly-coupled, $\Pmcr \ll 1$ limit (the CR fluid becomes entrained in the turbulent velocities and through the equilibrium flux equation, \autoref{eq:equilibrium_flux}, $\bm{F}_{\rm{c}} \approx \gamma E_{\rm{c}} \bm{v}$, hence $P_{F_{\rm{c}}}(k) \approx P_{v}(k)$). Overall, this makes $P_{F_{\rm{c}}}(k)$ a strong function of $\Pmcr$, directly changing the slope of $P_{F_{\rm{c}}}(k)$, which we explore in more detail via the correlation scales in the next section. The changes to the magnitude of $P_{F_{\rm{c}}}(k)$ with $\Pmcr$ is explored further in \autoref{sec:heating}.

\subsection{Correlation scales}
\begin{figure}
    \centering
    \includegraphics[width=0.98\linewidth]{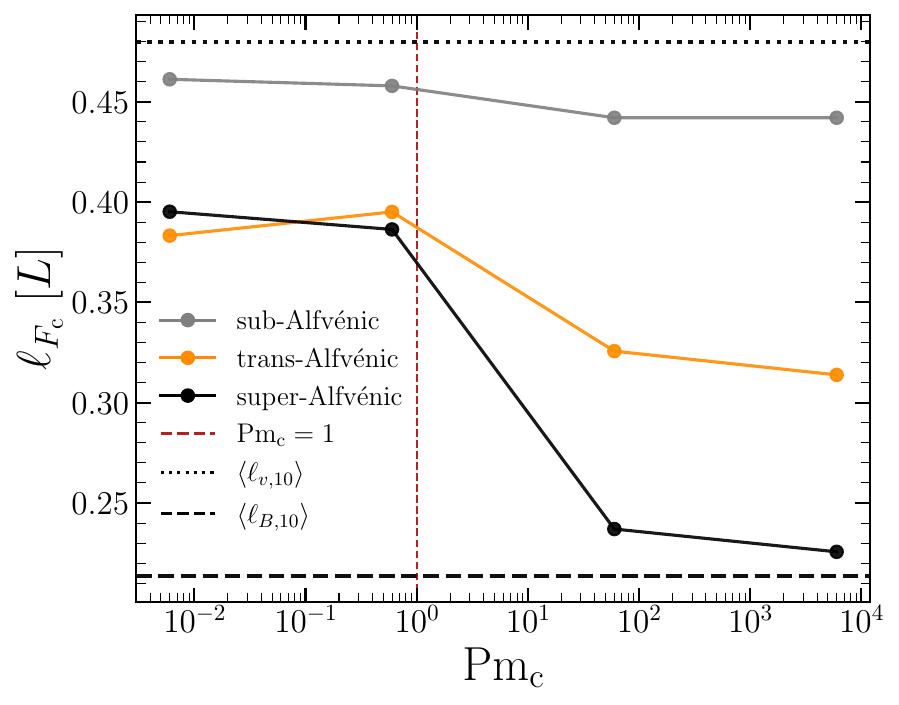}
    \caption{Measured correlation scales (\autoref{eq:corr}) for CR flux as a function of microphysical diffusion, $\Pmcr$. We show the isotropic correlation lengths with the color indicating the $\mao$ value of the simulation. We show a vertical line at $\Pmcr = 1 \iff \dcr = \dcrit$ and two horizontal lines indicating the mean values of $\ell_v$ and $\ell_B$ for the super-Alfv\'enic runs.}
    \label{fig:correlations}
\end{figure}

In \autoref{fig:correlations} we show the isotropic correlation lengths for $\bff_{\rm{c}}$ as a function of $\Pmcr$ for all simulations, as defined at the beginning of this section. The $\mao$ regime of the simulation is shown in colors, and we indicate $\Pmcr = 1 \iff \dcr = \dcrit$ with a red vertical line. Furthermore, we show the averaged values of the correlation lengths for velocity $\langle \ell_{v,10}\rangle$ and magnetic field $\langle \ell_{B,10} \rangle$ with the horizontal dashed lines in the bottom row from only super-Aflv\'enic runs, which span across the weakly-coupled to strongly-coupled regime, unlike the other simulations.

For the sub-Alfv\'enic simulations we see no significant change in the correlation lengths of any of the fields (reported in full in  \autoref{tab:trials}) as a function of $\Pmcr$, indicating a complete decoupling due to the high streaming speeds, low $\Pm_{\rm s}$, between CRs and the turbulence, as we observed previously in \autoref{sec:pdfs}. The super-Alfv\'enic simulations show that $\ell_{F_{\rm{c}}}$ is highly-dependent upon $\Pmcr$. Furthermore, and consistent with what we find for the spectra in the previous section, $\ell_{F_{\rm{c}}}$ is bounded between $\ell_{F_{\rm{c}}}\approx \langle \ell_{v,10} \rangle$ in the strongly-coupled, low-$\Pmcr$ limit, and $\ell_{F_{\rm{c}}} \approx \langle \ell_{B,10} \rangle$, in the weakly-coupled, high-$\Pmcr$ limit. In our trans-Alfv\'enic runs we see a weaker, yet still present effect of $\Pmcr$ on the value of $\ell_{F_{\rm{c}}}$ with the same general trend as in the super-Aflv\'enic runs.

In summary, we learn that in the trans-to-super-Alfv\'enic regime, the CR flux is imprinted with the correlation structure of the turbulent magnetic field if the CR-plasma coupling is weak (and since $\mao > 1$, dominated by diffusive fluxes) and the turbulent velocity when the CR-plasma coupling is strong. This is true not only on the outer-scale of the correlation, as we show in this sub-section, but also approximately for the entire flux spectrum itself, as we show in \autoref{fig:power-spectra}.

\section{Heating and reacceleration of the cosmic ray fluid}
\label{sec:heating}
From \autoref{fig:2d_pdf_energy_density} we observed that there is an increase in the $\left\langle E_{\rm{c}} \right\rangle$ as a function of the CR-plasma coupling strength. This is illustrated clearly in both the middle and right-most columns, which show the $E_{\rm{c}}$-$\rho$ PDFs for the trans- and super-Aflv\'enic plasmas, i.e., the plasmas that span the weakly-coupled to strongly-coupled regimes. A shift in $\left\langle E_{\rm{c}} \right\rangle$ is directly associated with the heating (or reacceleration) of the CR fluid, $Q_{\rm{c}}$, through the plasma itself.

Let us first derive the $\left\langle E_{\rm{c}} \right\rangle$ equation to explore $Q_{\rm{c}}$. In the stationary flux limit ($\partial_t \bff_{\rm{c}} \to 0$) \autoref{eqn:energy} and \autoref{eqn:flux} reduce to the one-moment description for a CR fluid,
\begin{equation}
\label{eq:one-mom}
    \frac{\partial E_{\rm{c}}}{\partial t}  + \nabla\cdot(E_{\rm{c}}[\bfv + \bfv_s]) = \underbrace{-P_{\rm{c}} \nabla \cdot (\bfv + \bfv_s)}_{\delta W} + \nabla \cdot \left(\frac{\mathbb{D}_c}{3} \cdot\nabla E_{\rm{c}}\right),
\end{equation}
where $\delta W = - P_{\rm{c}} \nabla \cdot (\bfv + \bfv_s) \approx -P_{\rm{c}}(\nabla\cdot\bfv + \bfb\cdot\nabla\rho/\rho^{3/2})$ is the work done by the plasma on the CR fluid. Integrating \label{eq:eq:one-mom} over space, we may express this as the heating rate for the CR fluid, $Q_{\rm c }$, 
\begin{equation}
\label{eq:heating}
    Q_{c} \equiv \frac{\partial \left\langle E_{\rm{c}} \right\rangle}{\partial t} = - \left\langle P_{\rm{c}} \nabla \cdot (\bfv + \bfv_s) \right\rangle,
\end{equation}
where $Q_{c} > 0$ is heating, $Q_{c} < 0$ cooling and we have have utilised that $\iiint \partial_iF_i dV = \oiint_{\mathcal{S}} F_i d\mathcal{S}_i = 0$ for periodic boundaries, getting rid of all of the conservative transport terms. We use this reduced equation to explore the evolution of $Q_{\rm{c}}$ in different CR-plasma limits in the following subsections.

\begin{figure}[h!]
    \centering
    \includegraphics[width=0.99\linewidth]{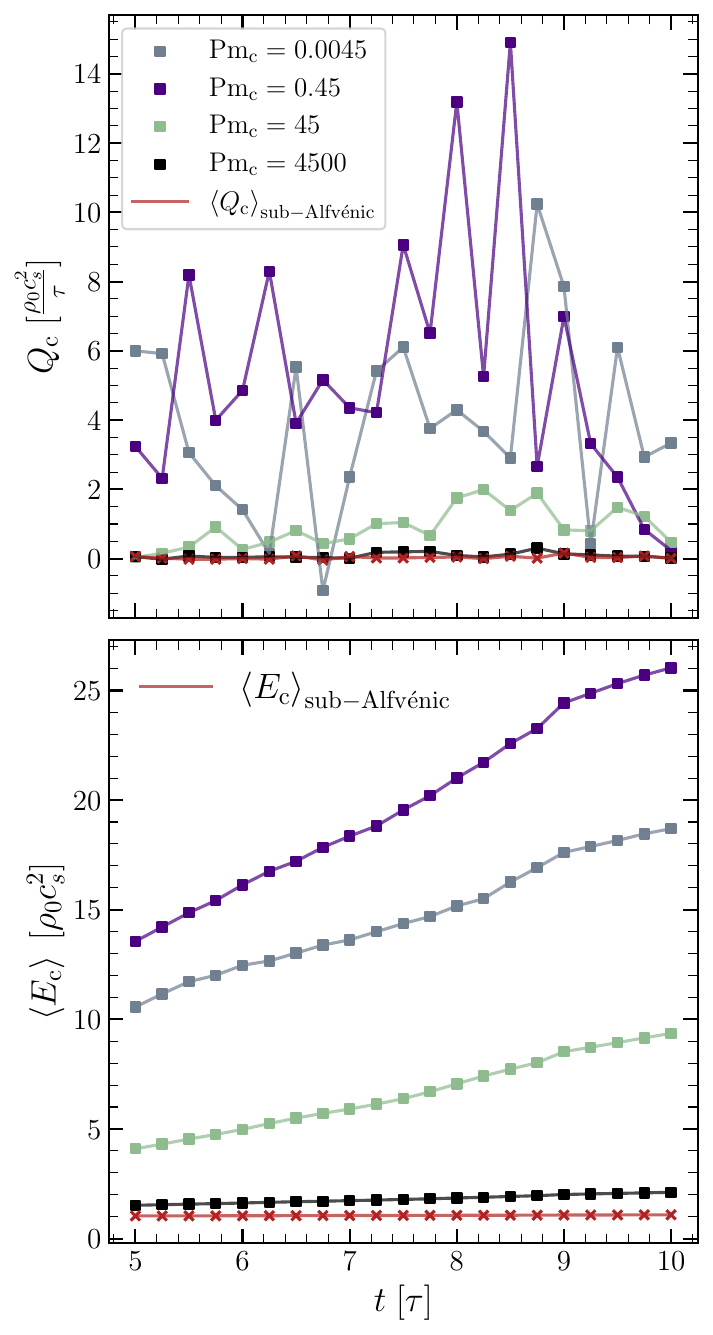}
    \caption{Cosmic ray heating rate, $Q_{\rm{c}}$ (\autoref{eq:heating}),  and $\left\langle E_{\rm{c}} \right\rangle$ as a function of time for $\mao\approx10$ simulations, and averaged $\mao \approx 0.75$ simulations (red). We show results for all $\Pmcr$ values indicated in color.}
    \label{fig:heating}
\end{figure}

To accompany our theoretical investigation of $Q_{\rm{c}}$, we plot $Q_{\rm{c}}$, following \autoref{eq:heating} and $\left\langle E_{\rm{c}} \right\rangle$ time evolution directly for each of the super-Alfv\'enic runs in the bottom panel of \autoref{fig:heating}, as well as the average over the sub-Alfv\'enic runs, which end up being exactly the same due to the decoupling of the CR fluid and the plasma. In general, we find an approximately linear increase in $E_{\rm{c}}$ as a function of time, with the slope depending on $\Pmcr$. 

\begin{figure*}
    \centering
    \includegraphics[width=0.98\linewidth]{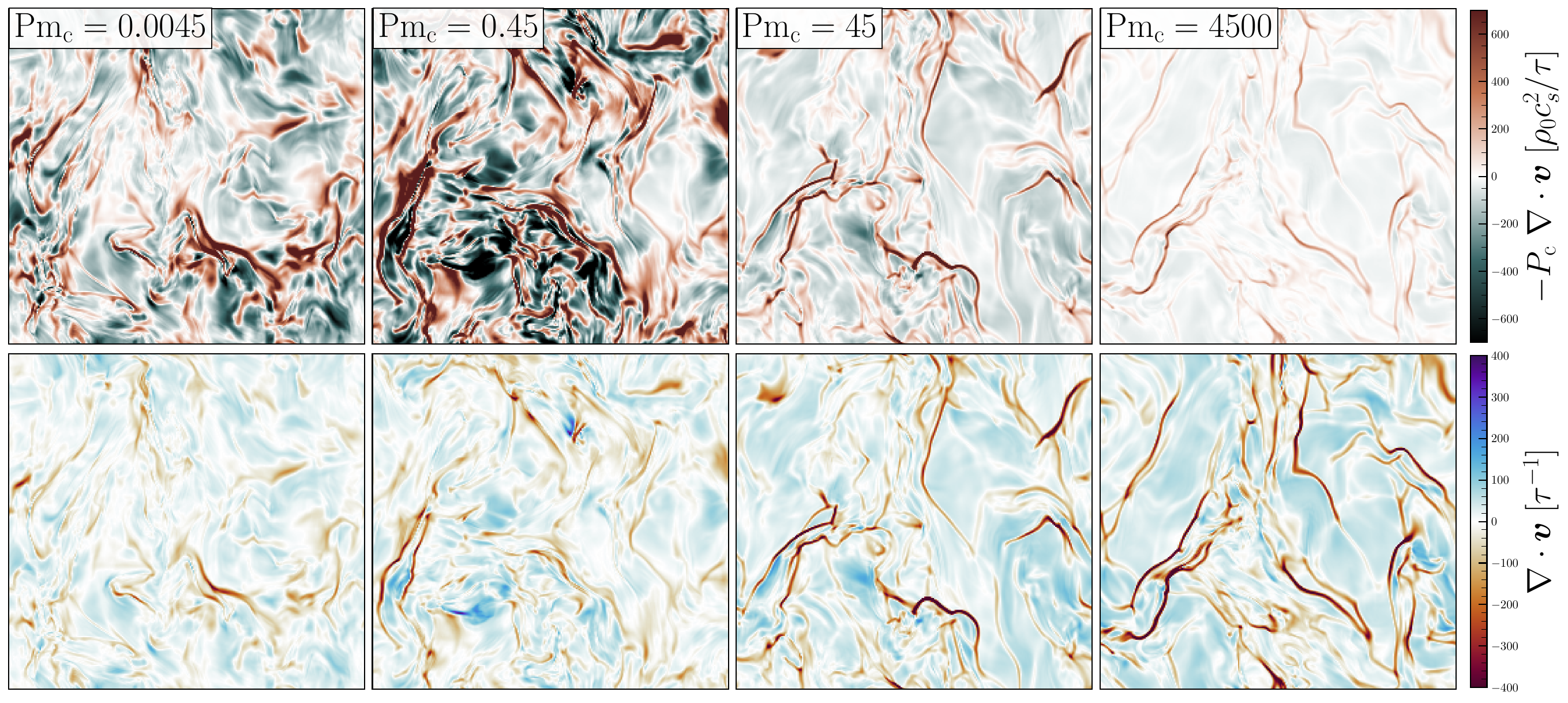}
    \caption{Two-dimensional slice visualizations through the perpendicular plane to $\bfb_0$, for the full suite of $\mao \approx 10$, super-Alfv\'enic simulations at $t=5\tau$. The columns show increasing levels of $\Pmcr$, and the rows showing $-P_{\rm{c}}\; \nabla \cdot \bfv$ and $\nabla \cdot \bfv$ in the top and bottom row respectively. The regions in the plasma that contribute to the the CR heating (\autoref{eq:heating}) are coherent structures where the flow is strongly-converging ($\nabla\cdot\bfv < 0$).}
    \label{fig:div-panels}
\end{figure*}

\subsection{Strong-diffusion limit}
Let us first consider the strong diffusion limit representing the bottom row of \autoref{fig:2d_pdf_energy_density}. Statistically, treating $\nabla \cdot (\bfv + \bfv_s)$ and $P_{\rm{c}}$ as random fields, $Q_{\rm{c}}$ becomes the covariance,
\begin{equation}
\label{eq:one-mom-cov}
    Q_{\rm{c}} = -\mathrm{cov}\left\{P_{\rm{c}}, \nabla \cdot (\bfv + \bfv_s)\right\},
\end{equation}
where $\mathrm{cov}\left\{ \hdots \right\}$ is the covariance operator. We note that we may exclude the $\langle P_{\rm{c}}\rangle\langle \nabla \cdot \bfv\rangle$ term\footnote{Note that $\langle XY \rangle = \langle (X-\langle X\rangle)(Y - \langle Y \rangle)\rangle + \langle X\rangle\langle Y\rangle = \rm{cov}\left\{X,Y \right\} + \langle X\rangle\langle Y\rangle$, for random variables $X$ and $Y$.} from the right-hand side of \autoref{eq:one-mom-cov} due to our periodic domain, $\langle \nabla \cdot (\bfv + \bfv_s)\rangle = 0$. Hence, we may relate the properties of $Q_{\rm{c}}$ to the properties of the PDFs we analyzed in \autoref{fig:2d_pdf_energy_density}. In this limit, and what is clear from \autoref{fig:2d_pdf_energy_density}, $E_{\rm{c}}$ and hence $P_{\rm{c}} = (\gamma_{\rm{c}} - 1)E_{\rm{c}}$ follows a uniform distribution, simply due to the short diffusion timescale of the CR fluid. This means that the right-hand-side (RHS) of \autoref{eq:one-mom-cov} $\mathrm{cov}\{P_{\rm{c}}, \nabla \cdot (\bfv + \bfv_s) \} \to 0$ and we expect to see no net gain in $E_{\rm{c}}$ throughout the simulation. This is demonstrated by the black line in \autoref{fig:heating}, showing that in the large $\Pmcr$, strong-diffusion limit, $\partial_t\left\langle E_{\rm{c}} \right\rangle = Q_{\rm{c}} \approx 0$, i.e.~$E_{\rm{c}} \approx E_{c,0} = 1$.

\subsection{Strong-$\bfb$ field limit}
The next case we consider is the strong $\bfb$-field limit where $\mao < 1$ and hence $\Pm_{\rm s} > 1$. In this regime $|\bfv_s| \gg |\bfv|$. Due to the very strong CR streaming fluxes, $\Pm_{\rm s} > 1$, like in the strong-diffusion case, $E_{\rm{c}}$ follows a uniform distribution throughout the domain. Hence, the RHS of \autoref{eq:one-mom} will tend towards zero. The strong-$\bfb$ field limit is indicated by the red line in \autoref{fig:heating}, which is the averaged $\left\langle E_{\rm{c}} \right\rangle$ as a function of time for all our sub-Alfv\'enic simulations. We see that weak-deviations from the strong-diffusion limit in the  super-Alfv\'enic simulations (black to teal lines) result in an increase in $Q_{\rm{c}}$ and hence growth in $\left\langle E_{\rm{c}} \right\rangle$. This is not true for the strong-$\bfb$ limit, which experience no $Q_{\rm{c}}$ for all $\Pmcr$ in this limit. In their CRMHD simulations \citet{bustard23} note that the hydrodynamical CR fluid is more efficiently heated in the diffusion dominated limits rather than in the strong-streaming limit, which appears consistent with what we show in \autoref{fig:heating} specifically for the super-Aflv\'enic regime.

\subsection{Weak $\bfb$-field, weak-diffusion limit}
The final case we consider is the weak $\bfb$-field ($\mao > 1; \Pm_{\rm s} < 1)$ slow diffusion ($\Pmcr < 1$) limit. In this case, $P_{\rm{c}}$ is not uniformly distributed, as shown in \autoref{fig:2d_pdf_energy_density}, and we have an increase in $\left\langle E_{\rm{c}} \right\rangle$ proportional to the negative covariance of $P_{\rm{c}}$ and $\nabla \cdot \bfv$ where we assume $|\bfv| \gg |\bfv_s|$ since the velocity field is super-Alfv\'enic ($\Pm_{\rm s} < 0.5$). 

\begin{figure}
    \centering
    \includegraphics[width=\linewidth]{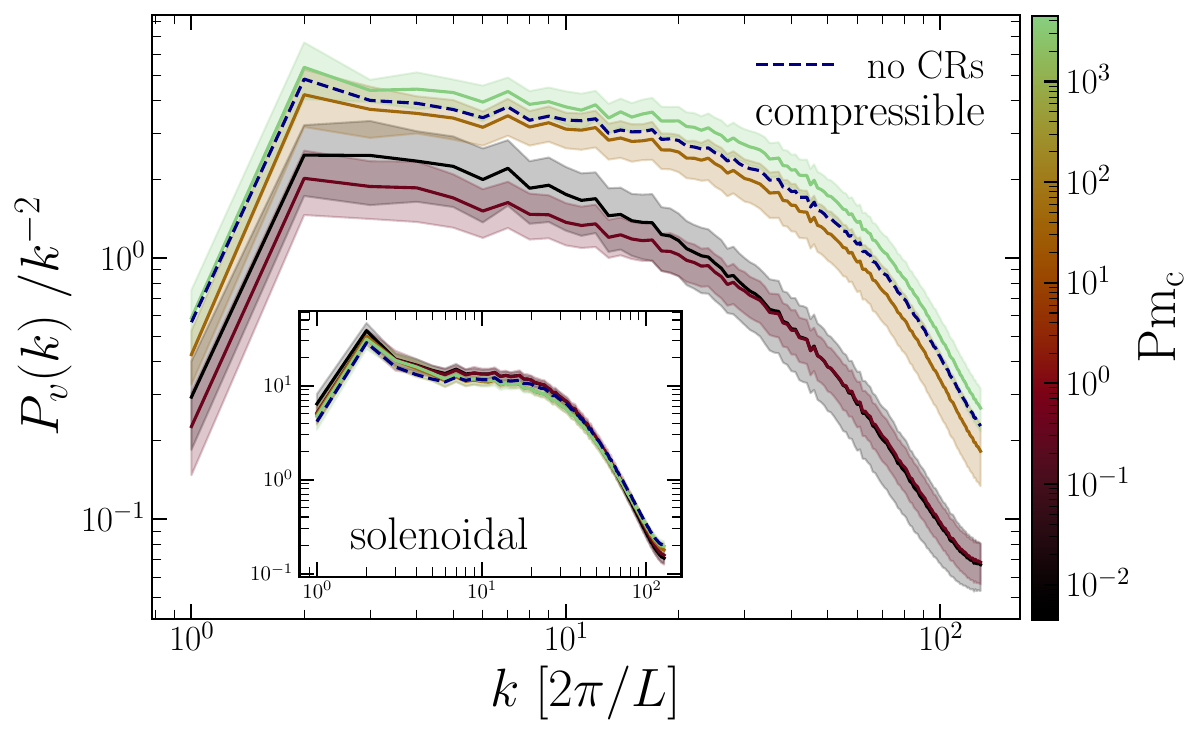}
    \caption{The same as in \autoref{fig:power-spectra} but for the velocity power spectra separated into compressible (velocity modes where $|\nabla\times\bfv|=0$) and solenoidal (velocities modes where $|\nabla\cdot\bfv|=0$) modes. We show the compressible mode spectrum in the main panel, and the solenoidal modes are shown in the inset. The compressible mode spectrum shows a strong dichotomy between the strongly-coupled (low-$\Pmcr$) and weakly-coupled (high-$\Pmcr$) regimes, reducing power on all $k$ when the CR fluid couples and becomes acoustically responsive. Unlike the compressible mode spectra, the solenoidal spectra remain invariant.}
    \label{fig:compressible_spectrum}
\end{figure}

We further note that the simulation gaining the largest energy (largest $Q_{\rm{c}}$ in \autoref{fig:heating}) is not the \emph{strongest-coupled} simulation \texttt{M4MA10D22} (gray in \autoref{fig:heating}), but instead our \texttt{M4MA10D24} simulation (purple in \autoref{fig:heating}). These results are mirrored in $P_{F_{\rm{c}}}(k)$ in the bottom-right panel of \autoref{fig:power-spectra}, where we see the magnitude of the spectra are the lowest for $\Pmcr \gg  1$ and the largest magnitude found in the simulation with the second lowest level of microphysical diffusion $\Pmcr=0.45$. This is initially counter-intuitive, as one would expect the largest $Q_{\rm{c}}$ to occur in the most strongly coupled CR-plasma regime. However,  whilst the coupling strength will allow for more efficient energy transfer, larger $Q_{\rm{c}}$, the strength of the coupling also increases the effective sound speed, reducing $\mathcal{M} \to\mathcal{M}_{\rm{eff}}$, as discussed in \autoref{sec:pdfs}. Due to the faster acoustic response, this smooths the $|\nabla\cdot\bfv|\neq 0$ regions, making the plasma less compressible and decreasing $Q_{\rm{c}}$, despite the strong coupling. This explains the lack of monotonic behavior in \autoref{fig:power-spectra} and \autoref{fig:heating}, where the plasma is transitioning between strongly-compressible (compressions are restored on the crossing time of $c_s$) and more weakly-compressible (compressions are restored on the crossing time of $c_{\rm eff} > c_s$) regimes.

In the bottom panel of \autoref{fig:heating} we see approximately linear growth across all $\Pmcr$ levels. Following arguments from \citet{Bustard2022}, if we expect CRs to gain energy via adiabatic compressions modified by diffusion (something akin to a second-order Fermi process) as they move through turbulent eddies, then it is expected that $dP_c/dt \sim dE_c/dt \sim \rm{const}$ in saturation which may explain the linear growth. In the coupled regime, \citet{bustard23} show the growth in $E_c$ primarily comes from extracting energy from the compressible turbulent modes, altering both the structure and magnitude of the compressive mode power spectrum. We explore this further in our simulations in the following paragraphs.

In \autoref{fig:div-panels} we visualize $-P_{\rm{c}}\nabla \cdot \bfv$ (top row;  non-spatial average of $Q_c$) and $\nabla \cdot \bfv$ (bottom row) for the $\mao \approx 10$ simulations. In the bottom panel, the plasma compressions $\nabla\cdot\bfv < 0$ are indicated with dark orange, and the rarefactions  $\nabla\cdot\bfv > 0$  in blue. Each column shows increasing levels of $\Pmcr$ from left-to-right. Firstly, we find that the compressions, and hence heating, is organized into highly-coherent $\nabla\cdot\bfv<0$ structures. As $\Pmcr$ is reduced (left panel) and the CR fluid more strongly-coupled, the coherent structures are diffused. This corresponds to a larger magnitude in $-P_{\rm{c}}\nabla \cdot \bfv$, specifically in the $\nabla\cdot\bfv < 0$ coherent regions, which decreases as we move towards the decoupled $\Pmcr > 1$ regime, in agreement with what we observed earlier in \autoref{fig:power-spectra} and \autoref{fig:heating}. 

Similar to \citet{bustard23}, in \autoref{fig:compressible_spectrum} we show the spectra for the Hodge-Helmholtz decomposition of the velocity field in the super-Alfv\'enic simulations. The compressible ($|\nabla\times\bfv|=0$) mode spectra are shown in the main panel and the solenoidal ($\nabla\cdot\bfv=0$) mode in the inset. We see that for $\Pmcr < 1$ there is a significant reduction in the magnitude of $P_v(k)$ across all $k$ modes, with the slope also becoming slightly steeper in the the weakly coupled and MHD regime. This is qualitatively consistent with \citet{bustard23}. This is perhaps indicative that the coherent structures in \autoref{fig:div-panels}, which occupy all of $k$ space, are being reduced in amplitude everywhere in the plasma, which is qualitatively consistent with what we observe in the bottom row of \autoref{fig:div-panels}.

 We see no significant effect on the solenoidal velocity spectrum, regardless of $\Pmcr$. This is different to \citet[][see Fig.~10]{bustard23} who find that the inclusion of a purely diffusion CR fluid alters the solenoidal mode spectrum significantly. Compared to \citet{bustard23}, we inject energy in both solenoidal and compressible modes (as opposed to just compressible). This means that solenoidal modes do not have to be sourced via compressible mode interactions, as in \citet{bustard23}, and always end up being energetically dominant, even with a 50:50 mix in the driving composition \citep{Federrath2010,Beattie2025_nature_astronomy}. Hence, if the solenoidal modes on outer-scale of the turbulence are being continuously replenished, we find that the spectra are quite robust to the coupling strength of the CR fluid, and that the modifications to the plasma are mostly confined to the compressible modes.

\section{Conclusions and key results}
\label{sec:conclusion}
In this study we explore the nature of cosmic ray (CR) and plasma coupling through a suite of simulations using a two-moment CRMHD fluid, with the evolution of anisotropic diffusion, and streaming fluxes. We study the coupling between the fluids across a variety of parallel microphysical diffusion coefficients, $\dcr$, and Alfv\'enic Mach numbers defined with respect to the mean magnetic field, $\mao$, revealing that the cosmic ray fluid and plasma are only coupled in a relatively small region of parameter space, constrained by both the streaming and the diffusion relative to the background turbulence. 

\begin{itemize}
    \item We non-dimensionalize the interaction coefficient introduced in \citet{2mom}, which describes the contributions from both the CR diffusion and the streaming, coupling them to the MHD equations. This allows us to define three Prandtl numbers, $\Pmcr$, $\Pm_{\rm s}$ and $\Pm_{\rm f}$ (\autoref{eq:Pms}). $\Pmcr$ is the ratio between microphysical diffusion and turbulent diffusion. $\Pm_{\rm s}$ is the ratio between streaming diffusion and turbulent diffusion. $\Pm_{\rm f}$ is the ratio between $\Pm_{\rm s}$ and $\Pmcr$, describing the relative contribution of diffusion and streaming fluxes in the interaction coefficient. For values of $\Pmcr \gg 1 \text{ and/or } \Pm_{\rm s} \gg 1$ the diffusive and streaming terms act on very short timescales compared to the background turbulence, hence indicating a strong decoupling between the two fluids. \\

    \item We make direct comparisons to work by \citet{commercon19} who, using a solely diffusive flux model, find that diffusion coefficients up to $\dcr \lesssim \dcrit$ result in CR-plasma coupling, allowing the CR fluid to modify the dynamics of the plasma, relatively independent of $\mao$. In our study, with the inclusion of CR streaming, we find the values of $\Pmcr$ at which the CR fluid is decoupled from the plasma depends strongly on $\mao$, hence $\Pm_{\rm s}$. Through a variety of statistics, we show that decoupling occurs at $\Pmcr \geq 1$ in all plasma regimes (e.g., strongly or weakly-diffusive CRs). \\
    
    \item For $\Pmcr \lesssim 1$, we find that for weakly-magnetized plasmas (trans and super-Alfv\'enic, but especially super-Alfv\'enic) the CRs and plasma become dynamically coupled. In particular the CRs imprint a mixture of the standard relativistic equation of state $P_{\rm{c}} \propto \rho^{4/3}$ and streaming equation of state $P_{\rm{c}} \propto \rho^{2/3}$, on the plasma (see \autoref{fig:2d_pdf_energy_density}). This acts to increase the effective sound speed of the plasma, in turn reducing the turbulent, sonic Mach number, and narrowing the width of the plasma density PDFs (see \autoref{fig:1Dpdf}), smoothing out over-densities and under-densities. This process leaves the correlation scale of the density invariant, and self-similarly reduces the power in all of the density modes in the density spectrum (see \autoref{fig:power-spectra}). \\

    \item We measure the power spectra and correlation lengths of the plasma density, magnetic field, velocity field and CR flux (see \autoref{fig:power-spectra} and \autoref{fig:correlations}). We find that the magnetic (which appear consistent with $P_B(k) \propto k^{-9/5}$ at high $k$) and velocity (which follow $P_v(k) \propto k^{-2}$ over a limited range of supersonic modes that are resolved) spectra are mostly invariant to changing how strongly the CR fluid couples to the plasma. Through a Hodge-Helmholtz decomposition we see that the compressible velocity modes are effected by the inclusion of a coupled CR fluid, with the magnitude of the compressible spectrum being reduced in the coupled regime. However, the solenoidal velocity spectrum, which makes up the bulk of the total energy of $\bfv$ in our simulation is invariant to $\Pmcr$. This provides tentative evidence that only the cascade physics of the solenoidal velocity modes (at the level of the spectral slopes), at least on the supersonic range of scales, does not change in the presence of a strongly-coupled CR fluid, with any CR heating coming from changes made purely to compressible modes. The plasma density and CR flux density fluctuations change significantly for different coupling strengths. We find that the fluctuations in $\bff_{\rm{c}}$ exhibit distinct regime change between the weak-diffusion (coupled) and strong-diffusion (decoupled) limits in the super-Alfv\'enic runs. We see the correlation scales $\ell_{F_{\rm{c}}} \lesssim \ell_{v}$ for $\Pmcr < 1$, while $\ell_{F_{\rm{c}}}\to \ell_{B}$ for $\Pmcr > 1$, i.e., the CR flux is bounded between the structure of the velocity and magnetic field in the two limits. The same trend is reflected directly in the spectra. \\

    \item In simulations with both weak magnetic fields, and weak levels of diffusion ($\Pmcr, \Pm_{\rm s} < 1$), we find $\left\langle E_{\rm{c}}\right\rangle$ increases linearly in time due to $-\langle P_{\rm{c}} \nabla\cdot \bfv\rangle$ heating. In \autoref{fig:div-panels} we show the structures heating the CR fluid are predominately plasma compressions ($\nabla\cdot\bfv < 0$) organized into coherent structures. As the coupling strength increases, the acoustic response from the CR fluid smooths the coherent structures, which results in the power spectrum of the compressible velocity modes decreasing across all $k$, but leaves the incompressible velocity modes unchanged. 
\end{itemize}

\subsection{Limitations of current  study}
In this study we present a detailed exploration of a range of the key aspects associated with CR fluid - MHD turbulence coupling. However, due in part to the particular decisions we have made there are a number of limitations in our study, which we list in detail below.

\subsubsection{Streaming at $v_A$ $(\chi = 1)$}
In the CR transport model we use streaming speeds set to $v_s = v_A$. While this is a reasonable assumption in highly-ionized plasmas where the ionized density $\rho_i \approx \rho$, this assumption does not hold in neutral regions.\footnote{We point out that in neutral rich phases of the ISM, CR streaming due to the CRSI may not be a valid assumption as ion-neutral damping may completely suppress the instability (e.g., Fig.~1 in \citealt{amato2018cosmic}).} The ion Alfv\'en velocity $v_{A,\rm{ion}}$ is a factor of $1/\sqrt{\chi}$ larger than $v_A$, where $\chi$ is the ionization fraction by mass. By mass, the warm and cold phases of the interstellar plasma can take on values of $\chi = 10^{-2}-10^{-8}$ which would increase the magnitude of $v_s$ by $10 - 10^4$ compared to the value used in our simulations \citep{draine2010physics,Ferriere2020_ISM_plasma_parameters,Krumholz2020CosmicGalaxies,Beattie2022_va_fluctuations}. This would result in high-$\Pm_{\rm{s}}\equiv \text{low-}\mao$ (using $v_{A,\rm{ion}}$ instead of $v_A$)  plasmas that would be dominated by significant levels of CR streaming. For example a plasma with $\mao=10$ and $\chi=10^{-4}$ would have the same streaming speed as a fully ionized plasma with $\mao=0.1$ (at least in terms of the mean magnetic field statistics; note that the total Alfv\'enic Mach number $\mathcal{M}_{A} = \sigma_v \sqrt{\rho}/(B_0+\delta B) \approx 4$ for the total magnetic field (turbulent, $\delta B$, and mean, $B_0$) in the $\mao=10$ regime; \citealt{Beattie2022_energybalance}). This is an important limitation, potentially constraining the CR fluid - plasma coupling to an even smaller region of parameter space in which we see coupling between the CR fluid and the plasma, e.g., $\Pmcr < 1$ and $\Pm_{\rm s} < 1$ 
where now $\Pm_{\rm s}$ is defined as $\sqrt{\chi}/\mao$. We therefore would expect decoupling for $\mao < 1/\sqrt{\chi}$ even in a zero-diffusion limit. 

\subsubsection{Limited grid resolution means limited plasma Reynolds numbers}
As stated in \autoref{sec:methods}, we run all simulations on a $256^3$ grid which results in effective plasma and magnetic Reynolds numbers between $\approx 500-1000$ for second-order spatial reconstructions \citep{Shivakumar2025_numerical_dissipation}. The Reynolds number of even a 10$\,\rm{pc}$ cold clump in the ISM could be as high as $\rm{Re}\sim 10^9$ \citep{Ferriere2020_ISM_plasma_parameters}. Our limited resolution is most noticeable in the power spectra in \autoref{fig:power-spectra} where the range of $k$ modes were not influenced by low-$k$ driving or high-$k$ numerical dissipation only spans a factor of a few. \citet{bustard23}, who use a grid resolution of $512^3$, find the inclusion of a diffusive CR fluid has significant impact on the shape of the kinetic energy spectrum at both large and small scales, i.e., that the turbulent cascade is modified by the additional CR pressure, but even those simulations are severely truncated, with a fully-developed cascade forming at grid resolutions above resolutions of $2,\!000^3$ for supersonic \citep{Federrath2013_universality_of_supersonic_trub,Beattie2025_nature_astronomy} and subsonic turbulence \citep{Fielding2023_plasmoids,Grete2023_dynamical_range}. We aim to probe the effect of a CR fluid in supersonic turbulence on the turbulent cascade in future work with larger grid resolutions and hence smaller (higher $k$) dissipation scales.

\subsubsection{Ideal MHD}
For all simulations in this paper, our plasma is evolved according to the ideal MHD equations. We do not capture any non-ideal effects, assuming that we probe the CRs and plasma on scales larger than the ion-neutral decoupling scale. This means we are not capturing effects such as non-linear Landau damping, ion-neutral damping, as well as other kinetic processes that have important effects on microphysical CR transport \citep{palenzuela2009, grassi19}.

\subsubsection{CR fluid approach}
It is also important to note that we are using a fluid approach to model CR transport, as opposed to the various Lagrangian approaches to this problem such as a test or passive tracer particle approach \citep{giacalone99,plotnikov2011,snodin2016global,wiener19,mertsch2020,criptic22} or via particle-in-cell methods \citep{baiPIC15,mignone18,bai2019magnetohydrodynamic,bambic2021,bai22,Ji22,sun23}. In general, these models capture the dynamics of individual CRs gyrating around magnetic field lines at close to the speed of light. In a fluid approach, we are limited to probe only scales larger than the mean free path $\lambda_{\rm{mfp}}$ of individual CRs. In other words, we assume we are simulating scales where there has been an isotropization of CR pitch angles ($\sim \lambda_{\rm{mfp}}$; \citealt{Krumholz2020CosmicGalaxies}) and we can adequately model CR transport with a streaming and diffusion approach. For GeV CRs in galactic environments $\lambda_{\rm{mfp}} \approx 0.1-10\,\rm{pc}$ \citep{wentzel74, garcia1987cosmic, yan2004cosmic, zweibel2013microphysics, Butsky2023, kempski24} hence we model only scales well beyond $\lambda_{\rm{mfp}}$.

\subsubsection{Periodic boundary conditions}
It is also important to note that we use periodic boundary conditions for all simulations. This has a potential effect on the coupling of streaming dominated regimes where in our simulation, CRs are not able to escape our domain. If we were to probe larger scales, allowing CRs to simultaneously escape and new CRs to be injected into the plasma, we may find CR-plasma coupling even in streaming dominated regimes. However, we note that the length-scale on which the coupling would be observed would likely be larger than $\ell_0$. \\

\section*{Acknowledgments}
We would like to thank Shaunak Modak, and Zach Hemler for their helpful discussions. M.~L.~S acknowledges funding from the Princeton University PhD fellowship. J.~R.~B. acknowledges the support from NSF Award~2206756. P.~K. was supported by the Lyman Spitzer, Jr. Fellowship at Princeton University. B.~C., Y.~D. and J.~R. acknowledge support by the Actions Thématiques ATPS, ATHE, ATCG and PCMI of CNRS/INSU PN Astro co-funded by CEA and CNES. This work was performed with the Princeton Research Computing resources at provided by Princeton University. 

\bibliography{Refs}

\begin{thebibliography}{}
\expandafter\ifx\csname natexlab\endcsname\relax\def\natexlab#1{#1}\fi
\providecommand{\url}[1]{\href{#1}{#1}}

\bibitem[{{Amato} \& {Blasi}(2018)}]{amato2018cosmic}
{Amato}, E., \& {Blasi}, P. 2018, Advances in Space Research, 62, 2731

\bibitem[{{Armillotta} {et~al.}(2021){Armillotta}, {Ostriker}, \& {Jiang}}]{armilotta2021}
{Armillotta}, L., {Ostriker}, E.~C., \& {Jiang}, Y.-F. 2021, \apj, 922, 11

\bibitem[{{Bai}(2022)}]{bai22}
{Bai}, X.-N. 2022, \apj, 928, 112

\bibitem[{{Bai} {et~al.}(2015){Bai}, {Caprioli}, {Sironi}, \& {Spitkovsky}}]{baiPIC15}
{Bai}, X.-N., {Caprioli}, D., {Sironi}, L., \& {Spitkovsky}, A. 2015, \apj, 809, 55

\bibitem[{Bai {et~al.}(2019)Bai, Ostriker, Plotnikov, \& Stone}]{bai2019magnetohydrodynamic}
Bai, X.-N., Ostriker, E.~C., Plotnikov, I., \& Stone, J.~M. 2019, The Astrophysical Journal, 876, 60

\bibitem[{{Bambic} {et~al.}(2021){Bambic}, {Bai}, \& {Ostriker}}]{bambic2021}
{Bambic}, C.~J., {Bai}, X.-N., \& {Ostriker}, E.~C. 2021, \apj, 920, 141

\bibitem[{{Beattie} \& {Federrath}(2020)}]{Beattie_2020_filaments_and_striations}
{Beattie}, J.~R., \& {Federrath}, C. 2020, \mnras, 492, 668

\bibitem[{Beattie {et~al.}(2025)Beattie, Federrath, Klessen, Cielo, \& Bhattacharjee}]{Beattie2025_nature_astronomy}
Beattie, J.~R., Federrath, C., Klessen, R.~S., Cielo, S., \& Bhattacharjee, A. 2025, Nature Astronomy, doi:10.1038/s41550-025-02551-5.
\newblock \url{https://doi.org/10.1038/s41550-025-02551-5}

\bibitem[{{Beattie} {et~al.}(2020){Beattie}, {Federrath}, \& {Seta}}]{beattie2020magnetic}
{Beattie}, J.~R., {Federrath}, C., \& {Seta}, A. 2020, \mnras, 498, 1593

\bibitem[{{Beattie} {et~al.}(2022{\natexlab{a}}){Beattie}, {Krumholz}, {Federrath}, {Sampson}, \& {Crocker}}]{Beattie2022_va_fluctuations}
{Beattie}, J.~R., {Krumholz}, M.~R., {Federrath}, C., {Sampson}, M., \& {Crocker}, R.~M. 2022{\natexlab{a}}, arXiv e-prints, arXiv:2203.13952

\bibitem[{{Beattie} {et~al.}(2022{\natexlab{b}}){Beattie}, {Krumholz}, {Skalidis}, {Federrath}, {Seta}, {Crocker}, {Mocz}, \& {Kriel}}]{Beattie2022_energybalance}
{Beattie}, J.~R., {Krumholz}, M.~R., {Skalidis}, R., {et~al.} 2022{\natexlab{b}}, \mnras, 515, 5267

\bibitem[{{Beattie} {et~al.}(2021){Beattie}, {Mocz}, {Federrath}, \& {Klessen}}]{beattie2021multishock}
{Beattie}, J.~R., {Mocz}, P., {Federrath}, C., \& {Klessen}, R.~S. 2021, \mnras, 504, 4354

\bibitem[{{Beattie} {et~al.}(2022{\natexlab{c}}){Beattie}, {Mocz}, {Federrath}, \& {Klessen}}]{Beattie2021_spdf}
---. 2022{\natexlab{c}}, \mnras, 517, 5003

\bibitem[{{Bieber} {et~al.}(1988){Bieber}, {Smith}, \& {Matthaeus}}]{bieber1988}
{Bieber}, J.~W., {Smith}, C.~W., \& {Matthaeus}, W.~H. 1988, \apj, 334, 470

\bibitem[{{Booth} {et~al.}(2013){Booth}, {Agertz}, {Kravtsov}, \& {Gnedin}}]{booth2013}
{Booth}, C.~M., {Agertz}, O., {Kravtsov}, A.~V., \& {Gnedin}, N.~Y. 2013, \apjl, 777, L16

\bibitem[{{Breitschwerdt} {et~al.}(1991){Breitschwerdt}, {McKenzie}, \& {Voelk}}]{oneMom}
{Breitschwerdt}, D., {McKenzie}, J.~F., \& {Voelk}, H.~J. 1991, \aap, 245, 79

\bibitem[{{Brucy} {et~al.}(2020){Brucy}, {Hennebelle}, {Bournaud}, \& {Colling}}]{brucy_driving}
{Brucy}, N., {Hennebelle}, P., {Bournaud}, F., \& {Colling}, C. 2020, \apjl, 896, L34

\bibitem[{Burgers(1948)}]{Burgers1948_turbulence}
Burgers, J. 1948, Advances in Applied Mechanics, 1, 171

\bibitem[{{Bustard} \& {Oh}(2022)}]{Bustard2022}
{Bustard}, C., \& {Oh}, S.~P. 2022, \apj, 941, 65

\bibitem[{{Bustard} \& {Oh}(2023)}]{bustard23}
---. 2023, \apj, 955, 64

\bibitem[{{Butsky} {et~al.}(2023){Butsky}, {Nakum}, {Ponnada}, {Hummels}, {Ji}, \& {Hopkins}}]{Butsky2023}
{Butsky}, I.~S., {Nakum}, S., {Ponnada}, S.~B., {et~al.} 2023, \mnras, 521, 2477

\bibitem[{{Caprioli} {et~al.}(2009){Caprioli}, {Blasi}, {Amato}, \& {Vietri}}]{caprioli2009}
{Caprioli}, D., {Blasi}, P., {Amato}, E., \& {Vietri}, M. 2009, \mnras, 395, 895

\bibitem[{{Cernetic} {et~al.}(2024){Cernetic}, {Springel}, {Guillet}, \& {Pakmor}}]{Cernetic2024_supersonic_turbulence_GPU}
{Cernetic}, M., {Springel}, V., {Guillet}, T., \& {Pakmor}, R. 2024, \mnras, 534, 1963

\bibitem[{{Cesarsky}(1980)}]{cesarsky80}
{Cesarsky}, C.~J. 1980, \araa, 18, 289

\bibitem[{{Commer{\c{c}}on} {et~al.}(2019){Commer{\c{c}}on}, {Marcowith}, \& {Dubois}}]{commercon19}
{Commer{\c{c}}on}, B., {Marcowith}, A., \& {Dubois}, Y. 2019, \aap, 622, A143

\bibitem[{{Crocker} {et~al.}(2021){Crocker}, {Krumholz}, \& {Thompson}}]{crocker2021cosmic}
{Crocker}, R.~M., {Krumholz}, M.~R., \& {Thompson}, T.~A. 2021, \mnras, 503, 2651

\bibitem[{{Dashyan} \& {Dubois}(2020)}]{Dashyan2020}
{Dashyan}, G., \& {Dubois}, Y. 2020, \aap, 638, A123

\bibitem[{Draine(2010)}]{draine2010physics}
Draine, B.~T. 2010, Physics of the interstellar and intergalactic medium,  Princeton University Press

\bibitem[{{Dubois} \& {Commer{\c{c}}on}(2016)}]{dubois16}
{Dubois}, Y., \& {Commer{\c{c}}on}, B. 2016, \aap, 585, A138

\bibitem[{{Dubois} {et~al.}(2019){Dubois}, {Commer{\c{c}}on}, {Marcowith}, \& {Brahimi}}]{Dubois2019}
{Dubois}, Y., {Commer{\c{c}}on}, B., {Marcowith}, A., \& {Brahimi}, L. 2019, \aap, 631, A121

\bibitem[{{Eswaran} \& {Pope}(1988)}]{eswaren_ornstein}
{Eswaran}, V., \& {Pope}, S.~B. 1988, Computers and Fluids, 16, 257

\bibitem[{{Farcy} {et~al.}(2022){Farcy}, {Rosdahl}, {Dubois}, {Blaizot}, \& {Martin-Alvarez}}]{farcy2022}
{Farcy}, M., {Rosdahl}, J., {Dubois}, Y., {Blaizot}, J., \& {Martin-Alvarez}, S. 2022, \mnras, 513, 5000

\bibitem[{{Farmer} \& {Goldreich}(2004)}]{Farmer04a}
{Farmer}, A.~J., \& {Goldreich}, P. 2004, \apj, 604, 671

\bibitem[{{Federrath}(2013)}]{Federrath2013_universality_of_supersonic_trub}
{Federrath}, C. 2013, \mnras, 436, 1245

\bibitem[{{Federrath} {et~al.}(2021){Federrath}, {Klessen}, {Iapichino}, \& {Beattie}}]{Federrath2021_sonic_scale}
{Federrath}, C., {Klessen}, R.~S., {Iapichino}, L., \& {Beattie}, J.~R. 2021, Nature Astronomy, 5, 365

\bibitem[{{Federrath} {et~al.}(2008){Federrath}, {Klessen}, \& {Schmidt}}]{Federrath2008}
{Federrath}, C., {Klessen}, R.~S., \& {Schmidt}, W. 2008, \apjl, 688, L79

\bibitem[{Federrath {et~al.}(2009)Federrath, Klessen, \& Schmidt}]{Federrath2009}
Federrath, C., Klessen, R.~S., \& Schmidt, W. 2009, \apj, 692, 364.
\newblock \url{http://arxiv.org/abs/0808.0605{\%}0Ahttp://dx.doi.org/10.1086/595280}

\bibitem[{Federrath {et~al.}(2010)Federrath, Roman-Duval, Klessen, Schmidt, \& {Mac Low}}]{Federrath2010}
Federrath, C., Roman-Duval, J., Klessen, R., Schmidt, W., \& {Mac Low}, M.~M. 2010, \aap, 512, arXiv:0905.1060.
\newblock \url{http://arxiv.org/abs/0905.1060{\%}0Ahttp://dx.doi.org/10.1051/0004-6361/200912437}

\bibitem[{{Federrath} {et~al.}(2010){Federrath}, {Roman-Duval}, {Klessen}, {Schmidt}, \& {Mac Low}}]{federrath2010comparing}
{Federrath}, C., {Roman-Duval}, J., {Klessen}, R.~S., {Schmidt}, W., \& {Mac Low}, M.~M. 2010, \aap, 512, A81

\bibitem[{{Ferri{\`e}re}(2020)}]{Ferriere2020_ISM_plasma_parameters}
{Ferri{\`e}re}, K. 2020, Plasma Physics and Controlled Fusion, 62, 014014

\bibitem[{{Fielding} {et~al.}(2023){Fielding}, {Ripperda}, \& {Philippov}}]{Fielding2023_plasmoids}
{Fielding}, D.~B., {Ripperda}, B., \& {Philippov}, A.~A. 2023, \apjl, 949, L5

\bibitem[{{Fromang} {et~al.}(2006){Fromang}, {Hennebelle}, \& {Teyssier}}]{fromang2006}
{Fromang}, S., {Hennebelle}, P., \& {Teyssier}, R. 2006, \aap, 457, 371

\bibitem[{{Gabici} {et~al.}(2010){Gabici}, {Casanova}, {Aharonian}, \& {Rowell}}]{Gabici2010_observational_D_W28}
{Gabici}, S., {Casanova}, S., {Aharonian}, F.~A., \& {Rowell}, G. 2010, in SF2A-2010: Proceedings of the Annual meeting of the French Society of Astronomy and Astrophysics, ed. S.~{Boissier}, M.~{Heydari-Malayeri}, R.~{Samadi}, \& D.~{Valls-Gabaud}, 313

\bibitem[{Garcia-Munoz {et~al.}(1987)Garcia-Munoz, Simpson, Guzik, Wefel, \& Margolis}]{garcia1987cosmic}
Garcia-Munoz, M., Simpson, J., Guzik, T., Wefel, J., \& Margolis, S. 1987, Astrophysical Journal Supplement Series (ISSN 0067-0049), vol. 64, May 1987, p. 269-304. Research supported by Louisiana State University and Washington University., 64, 269

\bibitem[{{Giacalone} \& {Jokipii}(1999)}]{giacalone99}
{Giacalone}, J., \& {Jokipii}, J.~R. 1999, \apj, 520, 204

\bibitem[{{Girichidis} {et~al.}(2022){Girichidis}, {Pfrommer}, {Pakmor}, \& {Springel}}]{girichidis22}
{Girichidis}, P., {Pfrommer}, C., {Pakmor}, R., \& {Springel}, V. 2022, \mnras, 510, 3917

\bibitem[{{Goldstein}(1976)}]{goldstein1976}
{Goldstein}, M.~L. 1976, \apj, 204, 900

\bibitem[{{Grassi} {et~al.}(2019){Grassi}, {Padovani}, {Ramsey}, {Galli}, {Vaytet}, {Ercolano}, \& {Haugb{\o}lle}}]{grassi19}
{Grassi}, T., {Padovani}, M., {Ramsey}, J.~P., {et~al.} 2019, \mnras, 484, 161

\bibitem[{{Grete} {et~al.}(2023){Grete}, {O'Shea}, \& {Beckwith}}]{Grete2023_dynamical_range}
{Grete}, P., {O'Shea}, B.~W., \& {Beckwith}, K. 2023, \apjl, 942, L34

\bibitem[{{Hanasz} {et~al.}(2013){Hanasz}, {Lesch}, {Naab}, {Gawryszczak}, {Kowalik}, \& {W{\'o}lta{\'n}ski}}]{hanasz2013}
{Hanasz}, M., {Lesch}, H., {Naab}, T., {et~al.} 2013, \apjl, 777, L38

\bibitem[{{Hanasz} {et~al.}(2021){Hanasz}, {Strong}, \& {Girichidis}}]{hanasz2021simulations}
{Hanasz}, M., {Strong}, A.~W., \& {Girichidis}, P. 2021, Living Reviews in Computational Astrophysics, 7, 2

\bibitem[{{Haverkorn}(2015)}]{Haverkorn2015_MW_magnetic_field_review}
{Haverkorn}, M. 2015, in Astrophysics and Space Science Library, Vol. 407, Magnetic Fields in Diffuse Media, ed. A.~{Lazarian}, E.~M. {de Gouveia Dal Pino}, \& C.~{Melioli}, 483

\bibitem[{{Heesen} {et~al.}(2023){Heesen}, {de Gasperin}, {Schulz}, {Basu}, {Beck}, {Br{\"u}ggen}, {Dettmar}, {Stein}, {Gajovi{\'c}}, {Tabatabaei}, \& {Reichherzer}}]{Heesen2023_D_observed_M51}
{Heesen}, V., {de Gasperin}, F., {Schulz}, S., {et~al.} 2023, \aap, 672, A21

\bibitem[{{Hopkins}(2013)}]{Hopkins2013_spdf}
{Hopkins}, P.~F. 2013, \mnras, 430, 1880

\bibitem[{{Hopkins} {et~al.}(2022{\natexlab{a}}){Hopkins}, {Butsky}, \& {Ji}}]{hopkins_2022_cR_model}
{Hopkins}, P.~F., {Butsky}, I.~S., \& {Ji}, S. 2022{\natexlab{a}}, arXiv e-prints, arXiv:2211.05811

\bibitem[{{Hopkins} {et~al.}(2021){Hopkins}, {Chan}, {Ji}, {Hummels}, {Kere{\v{s}}}, {Quataert}, \& {Faucher-Gigu{\`e}re}}]{hopkins2021cosmic}
{Hopkins}, P.~F., {Chan}, T.~K., {Ji}, S., {et~al.} 2021, \mnras, 501, 3640

\bibitem[{{Hopkins} {et~al.}(2022{\natexlab{b}}){Hopkins}, {Squire}, \& {Butsky}}]{Hopkins2022_two_moment_cRMHD_RSOL}
{Hopkins}, P.~F., {Squire}, J., \& {Butsky}, I.~S. 2022{\natexlab{b}}, \mnras, 509, 3779

\bibitem[{{Ji} {et~al.}(2022){Ji}, {Squire}, \& {Hopkins}}]{Ji22}
{Ji}, S., {Squire}, J., \& {Hopkins}, P.~F. 2022, \mnras, 513, 282

\bibitem[{{Jiang} \& {Oh}(2018)}]{2mom}
{Jiang}, Y.-F., \& {Oh}, S.~P. 2018, \apj, 854, 5

\bibitem[{{Jin} {et~al.}(2017){Jin}, {Salim}, {Federrath}, {Tasker}, {Habe}, \& {Kainulainen}}]{jin2017effective}
{Jin}, K., {Salim}, D.~M., {Federrath}, C., {et~al.} 2017, \mnras, 469, 383

\bibitem[{Jokipii \& Parker(1969)}]{jokipii1969cosmic}
Jokipii, J., \& Parker, E. 1969, The Astrophysical Journal, 155, 799

\bibitem[{{Jubelgas} {et~al.}(2008){Jubelgas}, {Springel}, {En{\ss}lin}, \& {Pfrommer}}]{jubelgas2008}
{Jubelgas}, M., {Springel}, V., {En{\ss}lin}, T., \& {Pfrommer}, C. 2008, \aap, 481, 33

\bibitem[{{Kempski} {et~al.}(2023){Kempski}, {Fielding}, {Quataert}, {Galishnikova}, {Kunz}, {Philippov}, \& {Ripperda}}]{kempski23}
{Kempski}, P., {Fielding}, D.~B., {Quataert}, E., {et~al.} 2023, \mnras, 525, 4985

\bibitem[{{Kempski} {et~al.}(2024){Kempski}, {Li}, {Fielding}, {Quataert}, {Phinney}, {Kunz}, {Jow}, \& {Philippov}}]{kempski24}
{Kempski}, P., {Li}, D., {Fielding}, D.~B., {et~al.} 2024, arXiv e-prints, arXiv:2412.03649

\bibitem[{{Kempski} \& {Quataert}(2020)}]{kempski20}
{Kempski}, P., \& {Quataert}, E. 2020, \mnras, 493, 1801

\bibitem[{{Kempski} \& {Quataert}(2022)}]{kempski}
---. 2022, \mnras, 514, 657

\bibitem[{{Kim} \& {Ryu}(2005)}]{Kim2005_density_spectrum}
{Kim}, J., \& {Ryu}, D. 2005, \apjl, 630, L45

\bibitem[{{Kriel} {et~al.}(2025){Kriel}, {Beattie}, {Federrath}, {Krumholz}, \& {Hew}}]{Kriel2025_curvature_and_dynamo}
{Kriel}, N., {Beattie}, J.~R., {Federrath}, C., {Krumholz}, M.~R., \& {Hew}, J. K.~J. 2025, \mnras, 537, 2602

\bibitem[{{Krumholz} {et~al.}(2022){Krumholz}, {Crocker}, \& {Sampson}}]{criptic22}
{Krumholz}, M.~R., {Crocker}, R.~M., \& {Sampson}, M.~L. 2022, \mnras, 517, 1355

\bibitem[{{Krumholz} {et~al.}(2020){Krumholz}, {Crocker}, {Xu}, {Lazarian}, {Rosevear}, \& {Bedwell-Wilson}}]{Krumholz2020CosmicGalaxies}
{Krumholz}, M.~R., {Crocker}, R.~M., {Xu}, S., {et~al.} 2020, \mnras, 493, 2817

\bibitem[{{Kulsrud} \& {Pearce}(1969)}]{Kulsrud69a}
{Kulsrud}, R., \& {Pearce}, W.~P. 1969, \apj, 156, 445

\bibitem[{Kulsrud(2005)}]{kulsrud2005plasma}
Kulsrud, R.~M. 2005, Plasma physics for astrophysics (Princeton University Press)

\bibitem[{{Larson}(1981)}]{Larson1981_LarsonsLaw}
{Larson}, R.~B. 1981, \mnras, 194, 809

\bibitem[{{Lazarian} \& {Xu}(2021)}]{laza21}
{Lazarian}, A., \& {Xu}, S. 2021, \apj, 923, 53

\bibitem[{{Lazarian} {et~al.}(2023){Lazarian}, {Xu}, \& {Hu}}]{laza23}
{Lazarian}, A., {Xu}, S., \& {Hu}, Y. 2023, Frontiers in Astronomy and Space Sciences, 10, 1154760

\bibitem[{{Lemoine}(2023)}]{lemoine23}
{Lemoine}, M. 2023, Journal of Plasma Physics, 89, 175890501

\bibitem[{{Lerche}(1967)}]{Lerche67a}
{Lerche}, I. 1967, \apj, 147, 689

\bibitem[{{Mao} \& {Ostriker}(2018)}]{mao2018galactic}
{Mao}, S.~A., \& {Ostriker}, E.~C. 2018, \apj, 854, 89

\bibitem[{{Martin-Alvarez} {et~al.}(2023){Martin-Alvarez}, {Sijacki}, {Haehnelt}, {Farcy}, {Dubois}, {Belokurov}, {Rosdahl}, \& {Lopez-Rodriguez}}]{MartinAlvarez2023_cr_galaxy_models}
{Martin-Alvarez}, S., {Sijacki}, D., {Haehnelt}, M.~G., {et~al.} 2023, \mnras, 525, 3806

\bibitem[{{Mertsch}(2020)}]{mertsch2020}
{Mertsch}, P. 2020, \apss, 365, 135

\bibitem[{{Mignone} {et~al.}(2018){Mignone}, {Bodo}, {Vaidya}, \& {Mattia}}]{mignone18}
{Mignone}, A., {Bodo}, G., {Vaidya}, B., \& {Mattia}, G. 2018, \apj, 859, 13

\bibitem[{{Mocz} \& {Burkhart}(2019)}]{Mocz2019_markov_model}
{Mocz}, P., \& {Burkhart}, B. 2019, \apjl, 884, L35

\bibitem[{Molina {et~al.}(2012)Molina, Glover, Federrath, \& Klessen}]{Molina2012_dens_var}
Molina, F.~Z., Glover, S. C.~O., Federrath, C., \& Klessen, R.~S. 2012, \mnras, 423, 2680.
\newblock \url{https://doi.org/10.1111/j.1365-2966.2012.21075.x}

\bibitem[{{Nam} {et~al.}(2021){Nam}, {Federrath}, \& {Krumholz}}]{Nam2021_power_law_driving}
{Nam}, D.~G., {Federrath}, C., \& {Krumholz}, M.~R. 2021, \mnras, 503, 1138

\bibitem[{{Nolan} {et~al.}(2015){Nolan}, {Federrath}, \& {Sutherland}}]{Nolan2015_density_variance}
{Nolan}, C.~A., {Federrath}, C., \& {Sutherland}, R.~S. 2015, \mnras, 451, 1380

\bibitem[{{Orlando} {et~al.}(2012){Orlando}, {Bocchino}, {Miceli}, {Petruk}, \& {Pumo}}]{orlando2012}
{Orlando}, S., {Bocchino}, F., {Miceli}, M., {Petruk}, O., \& {Pumo}, M.~L. 2012, \apj, 749, 156

\bibitem[{{Padovani} {et~al.}(2009){Padovani}, {Galli}, \& {Glassgold}}]{padovani2009cosmic}
{Padovani}, M., {Galli}, D., \& {Glassgold}, A.~E. 2009, \aap, 501, 619

\bibitem[{{Palenzuela} {et~al.}(2009){Palenzuela}, {Lehner}, {Reula}, \& {Rezzolla}}]{palenzuela2009}
{Palenzuela}, C., {Lehner}, L., {Reula}, O., \& {Rezzolla}, L. 2009, \mnras, 394, 1727

\bibitem[{{Pfrommer} {et~al.}(2017){Pfrommer}, {Pakmor}, {Schaal}, {Simpson}, \& {Springel}}]{pfrommer17}
{Pfrommer}, C., {Pakmor}, R., {Schaal}, K., {Simpson}, C.~M., \& {Springel}, V. 2017, \mnras, 465, 4500

\bibitem[{{Plotnikov} {et~al.}(2011){Plotnikov}, {Pelletier}, \& {Lemoine}}]{plotnikov2011}
{Plotnikov}, I., {Pelletier}, G., \& {Lemoine}, M. 2011, \aap, 532, A68

\bibitem[{{Quataert} {et~al.}(2022){Quataert}, {Jiang}, \& {Thompson}}]{quataert22}
{Quataert}, E., {Jiang}, Y.-F., \& {Thompson}, T.~A. 2022, \mnras, 510, 920

\bibitem[{{Recchia} {et~al.}(2016){Recchia}, {Blasi}, \& {Morlino}}]{recchia16}
{Recchia}, S., {Blasi}, P., \& {Morlino}, G. 2016, \mnras, 462, 4227

\bibitem[{{Reichherzer} {et~al.}(2020){Reichherzer}, {Becker Tjus}, {Zweibel}, {Merten}, \& {Pueschel}}]{reichherzer2020turbulence}
{Reichherzer}, P., {Becker Tjus}, J., {Zweibel}, E.~G., {Merten}, L., \& {Pueschel}, M.~J. 2020, \mnras, 498, 5051

\bibitem[{{Reichherzer} {et~al.}(2022){Reichherzer}, {Becker Tjus}, {Zweibel}, {Merten}, \& {Pueschel}}]{reichherzer22}
---. 2022, \mnras, 514, 2658

\bibitem[{{Robertson} \& {Goldreich}(2018)}]{Robertson2018_dense_regions}
{Robertson}, B., \& {Goldreich}, P. 2018, \apj, 854, 88

\bibitem[{{Rodr{\'\i}guez Montero} {et~al.}(2024){Rodr{\'\i}guez Montero}, {Martin-Alvarez}, {Slyz}, {Devriendt}, {Dubois}, \& {Sijacki}}]{montero24}
{Rodr{\'\i}guez Montero}, F., {Martin-Alvarez}, S., {Slyz}, A., {et~al.} 2024, \mnras, 530, 3617

\bibitem[{{Ruszkowski} \& {Pfrommer}(2023)}]{Ruszkowski2023_cr_review}
{Ruszkowski}, M., \& {Pfrommer}, C. 2023, \aapr, 31, 4

\bibitem[{{Salem} {et~al.}(2014){Salem}, {Bryan}, \& {Hummels}}]{salem2014cosmological}
{Salem}, M., {Bryan}, G.~L., \& {Hummels}, C. 2014, \apjl, 797, L18

\bibitem[{{Sampson} {et~al.}(2023){Sampson}, {Beattie}, {Krumholz}, {Crocker}, {Federrath}, \& {Seta}}]{sampson_2022_turb}
{Sampson}, M.~L., {Beattie}, J.~R., {Krumholz}, M.~R., {et~al.} 2023, \mnras, 519, 1503

\bibitem[{{Schekochihin}(2022)}]{Schekochihin2022_a_bias_review}
{Schekochihin}, A.~A. 2022, Journal of Plasma Physics, 88, 155880501

\bibitem[{{Schmidt} {et~al.}(2009){Schmidt}, {Federrath}, {Hupp}, {Kern}, \& {Niemeyer}}]{schmidt2009numerical}
{Schmidt}, W., {Federrath}, C., {Hupp}, M., {Kern}, S., \& {Niemeyer}, J.~C. 2009, \aap, 494, 127

\bibitem[{{Shalchi} {et~al.}(2004){Shalchi}, {Bieber}, {Matthaeus}, \& {Qin}}]{shalchi2004nonlinear}
{Shalchi}, A., {Bieber}, J.~W., {Matthaeus}, W.~H., \& {Qin}, G. 2004, \apj, 616, 617

\bibitem[{{Shalchi} {et~al.}(2009){Shalchi}, {Skoda}, {Tautz}, \& {Schlickeiser}}]{shalchi2009analytical}
{Shalchi}, A., {Skoda}, T., {Tautz}, R.~C., \& {Schlickeiser}, R. 2009, \aap, 507, 589

\bibitem[{{Sharma} {et~al.}(2010){Sharma}, {Colella}, \& {Martin}}]{sharma2009}
{Sharma}, P., {Colella}, P., \& {Martin}, D.~F. 2010, SIAM Journal on Scientific Computing, 32, 3564

\bibitem[{{Shivakumar} \& {Federrath}(2025)}]{Shivakumar2025_numerical_dissipation}
{Shivakumar}, L.~M., \& {Federrath}, C. 2025, \mnras, 537, 2961

\bibitem[{{Simpson} {et~al.}(2016){Simpson}, {Pakmor}, {Marinacci}, {Pfrommer}, {Springel}, {Glover}, {Clark}, \& {Smith}}]{simpson2016role}
{Simpson}, C.~M., {Pakmor}, R., {Marinacci}, F., {et~al.} 2016, \apjl, 827, L29

\bibitem[{{Skilling}(1971)}]{Skilling71a}
{Skilling}, J. 1971, \apj, 170, 265

\bibitem[{{Skilling}(1975)}]{skilling1975cosmic}
---. 1975, \mnras, 172, 557

\bibitem[{{Snodin} {et~al.}(2016){Snodin}, {Shukurov}, {Sarson}, {Bushby}, \& {Rodrigues}}]{snodin2016global}
{Snodin}, A.~P., {Shukurov}, A., {Sarson}, G.~R., {Bushby}, P.~J., \& {Rodrigues}, L.~F.~S. 2016, \mnras, 457, 3975

\bibitem[{{Squire} \& {Hopkins}(2017)}]{Squire2017_s_pdf}
{Squire}, J., \& {Hopkins}, P.~F. 2017, \mnras, 471, 3753

\bibitem[{{Stepanov} {et~al.}(2014){Stepanov}, {Shukurov}, {Fletcher}, {Beck}, {La Porta}, \& {Tabatabaei}}]{stepanov2014}
{Stepanov}, R., {Shukurov}, A., {Fletcher}, A., {et~al.} 2014, \mnras, 437, 2201

\bibitem[{{Sun} \& {Bai}(2023)}]{sun23}
{Sun}, X., \& {Bai}, X.-N. 2023, \mnras, 523, 3328

\bibitem[{{Teyssier}(2002)}]{RAMSES}
{Teyssier}, R. 2002, \aap, 385, 337

\bibitem[{{Teyssier} {et~al.}(2006){Teyssier}, {Fromang}, \& {Dormy}}]{teyssier2006}
{Teyssier}, R., {Fromang}, S., \& {Dormy}, E. 2006, Journal of Computational Physics, 218, 44

\bibitem[{{Thomas} \& {Pfrommer}(2019)}]{thomas_19}
{Thomas}, T., \& {Pfrommer}, C. 2019, \mnras, 485, 2977

\bibitem[{{Thomas} \& {Pfrommer}(2022)}]{thomas22}
---. 2022, \mnras, 509, 4803

\bibitem[{{Thomas} {et~al.}(2021){Thomas}, {Pfrommer}, \& {Pakmor}}]{thomas21}
{Thomas}, T., {Pfrommer}, C., \& {Pakmor}, R. 2021, \mnras, 503, 2242

\bibitem[{{Thomas} {et~al.}(2023){Thomas}, {Pfrommer}, \& {Pakmor}}]{thomas23}
---. 2023, \mnras, 521, 3023

\bibitem[{{Uhlig} {et~al.}(2012){Uhlig}, {Pfrommer}, {Sharma}, {Nath}, {En{\ss}lin}, \& {Springel}}]{uhlig2012}
{Uhlig}, M., {Pfrommer}, C., {Sharma}, M., {et~al.} 2012, \mnras, 423, 2374

\bibitem[{{Wentzel}(1974)}]{wentzel74}
{Wentzel}, D.~G. 1974, \araa, 12, 71

\bibitem[{{Werhahn} {et~al.}(2021){Werhahn}, {Pfrommer}, {Girichidis}, \& {Winner}}]{werhahn2021}
{Werhahn}, M., {Pfrommer}, C., {Girichidis}, P., \& {Winner}, G. 2021, \mnras, 505, 3295

\bibitem[{{Wiener} {et~al.}(2017{\natexlab{a}}){Wiener}, {Oh}, \& {Zweibel}}]{wiener2017b}
{Wiener}, J., {Oh}, S.~P., \& {Zweibel}, E.~G. 2017{\natexlab{a}}, \mnras, 467, 646

\bibitem[{{Wiener} {et~al.}(2017{\natexlab{b}}){Wiener}, {Pfrommer}, \& {Oh}}]{wiener2017}
{Wiener}, J., {Pfrommer}, C., \& {Oh}, S.~P. 2017{\natexlab{b}}, \mnras, 467, 906

\bibitem[{{Wiener} {et~al.}(2019){Wiener}, {Zweibel}, \& {Ruszkowski}}]{wiener19}
{Wiener}, J., {Zweibel}, E.~G., \& {Ruszkowski}, M. 2019, \mnras, 489, 205

\bibitem[{{Xu} \& {Lazarian}(2022)}]{Xu2022_streaming}
{Xu}, S., \& {Lazarian}, A. 2022, \apj, 927, 94

\bibitem[{{Yan} \& {Lazarian}(2004)}]{yan2004cosmic}
{Yan}, H., \& {Lazarian}, A. 2004, \apj, 614, 757

\bibitem[{{Yuen} \& {Lazarian}(2020)}]{Yuen2020_curvature}
{Yuen}, K.~H., \& {Lazarian}, A. 2020, \apj, 898, 66

\bibitem[{{Zhdankin} {et~al.}(2017){Zhdankin}, {Boldyrev}, \& {Mason}}]{Zhdankin2017_large_scale_field_e_flux}
{Zhdankin}, V., {Boldyrev}, S., \& {Mason}, J. 2017, \mnras, 468, 4025

\bibitem[{Zweibel(2013)}]{zweibel2013microphysics}
Zweibel, E.~G. 2013, Physics of Plasmas, 20, 055501

\bibitem[{{Zweibel}(2017)}]{zweibel2017basis}
{Zweibel}, E.~G. 2017, Physics of Plasmas, 24, 055402

\end{thebibliography}

\appendix

\section{Dimensionless CR flux equation}\label{appendix:flux_equation}
Let us consider the dimensionless form of \autoref{eqn:flux}, where we scale each term by the natural units of system: turbulent velocities, $\sigma_{\bm{v}}$, correlation length scales, $\ell_0$, and mean plasma densities, $\rho_0$, and magnetic fields, $B_0$, i.e., 
\begin{align}
& \hat{\bm{x}} = \frac{\bm{x}}{\ell_0}, \quad \hat{t} = \frac{t}{\tau}, \quad \hat{\nabla} = \ell_0 \nabla, \quad \hat{\rho} = \frac{\rho}{\rho_0}, \quad \hat{\bm{v}} = \frac{\bm{v}}{\sigma_v}, \quad \hat{E}_c = \frac{E_{\rm{c}}}{\rho_0 \sigma_v^2}, \quad \hat{P}_c = \frac{P_{\rm{c}}}{\rho_0 \sigma_v^2}, \quad \hat{\bm{F}}_c = \frac{\bm{F}_{\rm{c}}}{\rho_0 \sigma_v^3}, \quad\hat{\bm{B}} = \frac{\bm{B}}{B_0},
\end{align}
noting that $\tau = \ell / \sigma_{\bm{v}}$ and $v_{A0} = B_0 / \sqrt{\rho_0}$. Let us rewrite \autoref{eqn:flux} with this scaling, 
\begin{align}
    \epsilon^2 \frac{\rho_0 \sigma_{\bm{v}}^2}{\ell_0} \frac{\partial \hat{\bm{F}}_c}{\partial \hat{t}} +
\frac{\rho_0 \sigma_v^2}{\ell_0} \hat{\nabla} \cdot \hat{\mathbb{P}}_c =-(\gamma_c-1)\bm{\sigma_c} \cdot \left[
\rho_0 \sigma_v^3 \hat{\bm{F}}_c - \rho_0 \sigma_{\bm{v}}^3 \hat{\bm{v}} (1 + 1/3)\hat{E}_c
\right]
\end{align}
where $\epsilon^2 = \sigma_{\bm v}^2 / V_m^2 \ll 1$. Now let us consider the interaction coefficient itself, which is the only remaining dimensional term,
\begin{align}
    \sigma_{\bm{v}}\ell_0 /\bm{\sigma_c}= \hat{\bm{\sigma}}_c^{-1} = \frac{\mathbb{D}_c}{\sigma_{\bm v}\ell_0} +\  \frac{1}{\mao}\frac{1}{\sqrt{\hat{\rho}}}\frac{\hat{\bfb} \otimes \hat{\bfb}}{\|\hat{\bfb} \cdot \hat{\bnab} \hat{P_{\rm{c}}}\|}  \cdot(\hat{\mathbb{E}}_c + \hat{\mathbb{P}}_c),
\end{align}
where, as we describe in the main text, $\dcrit = \sigma_{\bm{v}}\ell_0$, hence,
\begin{align}
        \hat{\bm{\sigma}}_c^{-1} = \frac{\mathbb{D}_c}{\dcrit} +\  \frac{\ds}{\dcrit}\frac{1}{\sqrt{\hat{\rho}}}\frac{\hat{\bfb} \otimes \hat{\bfb}}{\|\hat{\bfb} \cdot \hat{\bnab} \hat{P_{\rm{c}}}\|}  \cdot(\hat{\mathbb{E}}_c + \hat{\mathbb{P}}_c) \iff \hat{\bm{\sigma}}_c^{-1} = \Pmcr +\  \Pm_{\rm s}\frac{1}{\sqrt{\hat{\rho}}}\frac{\hat{\bfb} \otimes \hat{\bfb}}{\|\hat{\bfb} \cdot \hat{\bnab} \hat{P_{\rm{c}}}\|}  \cdot(\hat{\mathbb{E}}_c + \hat{\mathbb{P}}_c) .
\end{align}
We immediately see, in this non-dimensionalization, the diffusion term scales like $\Pmcr = \dcr/\dcrit$, and the streaming term like $\Pm_{\rm s} = \ds/\dcrit = 1/\mao$. Putting this back into the dimensionless flux equation, we find that we get a perfectly dimensionless equation, with all dimensionless numbers contained within $\bm{\sigma}_c$,
\begin{align}
    \epsilon^2 \frac{\partial \hat{\bm{F}}_c}{\partial \hat{t}} +
\hat{\nabla} \cdot \mathbb{P}_c =-(\gamma_c-1)\hat{\bm{\sigma_c}} \cdot \left[ \hat{\bm{F}}_c -\hat{\bm{v}} (1 + 1/3)\hat{E}_c
\right],
\end{align}
hence, it is the dimensionless form of $\bm{\sigma}_c$ that matters the most for the determining the different flux regimes.

\end{document}